\documentclass[12pt,twoside]{article}
\usepackage{float}
\usepackage{amsfonts}
\usepackage{amssymb}
\usepackage{amsmath}
\usepackage{amsthm}
\usepackage{mathrsfs}	
\usepackage{stmaryrd}
\usepackage{url}
\usepackage{listings}
\usepackage{color}
\usepackage{multicol}
\usepackage{graphicx}
\usepackage{wrapfig}
\usepackage{pgf}
\usepackage[english]{babel}
\usepackage{tikz}
\usetikzlibrary{matrix}
\usepackage{pgf}
\usetikzlibrary{arrows,automata}
\usepackage[ruled,vlined,commentsnumbered]{algorithm2e}
\tikzstyle{every picture}+=[remember picture]
\usepackage{setspace}
\usepackage{pbox}
\usetikzlibrary{fit,matrix}
\usepackage{datetime}
\newdate{date}{10}{07}{2014}
\usepackage{float}

\tikzset{%
  highlight/.style={rectangle,rounded corners,fill=#1!25,draw=#1!50,fill opacity=0.0,very thick,inner sep=0pt}
}
\newcommand{\tikzmark}[2]{\tikz[overlay,remember picture,baseline=(#1.base)] \node (#1) {#2};}

\newcommand{\Highlight}[3]{%
    \tikz[overlay,remember picture]{
    \node[highlight=#3,fit=(#1.north west) (#2.south east)]  {};}
}

\newtheorem{prop}{Proposition}[section]
\newtheorem{LM}{Lemma}[section]
\newtheorem{thm}{Theorem}[section]
\newtheorem{df}{Definition}[section]
\newtheorem{cor}{Corollary}[section]

\newtheorem{ex}{Example}[section]

\newcommand{\R}{\mathbb{R}}
\renewcommand{\l}{\lambda}

\newcommand{\Q}{\mathbb{Q}}
\renewcommand{\H}{\mathbb{H}}

\newcommand{\A}{\textbf{A}}

\newcommand{\G}{\mathbb{G}}

\renewcommand{\b}{\textbf{b}}
\newcommand{\F}{\mathbf{F}}
\renewcommand{\c}{\textbf{c}}

\newcommand{\N}{\mathbb{N}}
\newcommand{\Z}{\mathbb{Z}}

\renewcommand{\i}{\textbf{id\_E}}
\newcommand{\1}{\mathbf{1}}

\renewcommand{\a}{\textbf{a}}
\newcommand{\f}{\textbf{f}}
\newcommand{\x}{\textbf{x}}

\begin{document}%

\lstdefinestyle{customc}{
  belowcaptionskip=1\baselineskip,
  breaklines=true,
  frame=L,
  xleftmargin=\parindent,
  language=C,
  showstringspaces=false,
  basicstyle=\footnotesize\ttfamily,
  keywordstyle=\bfseries\color{green!40!black},
  commentstyle=\itshape\color{purple!40!black},
  identifierstyle=\color{blue},
  stringstyle=\color{orange},}

\lstset{escapechar=@,style=customc}

\title{Generating Asymptotically Non-Terminating Initial Values for Linear Programs}

\author{Rachid Rebiha\thanks{Instituto  de Computacao, Universidade Estadual  de Campinas, 13081\-970  Campinas,  SP.  Pesquisa desenvolvida com suporte financeiro da FAPESP, processos 2011\/08947\-1 e FAPESP BEPE 2013\/04734\-9 } \and  Arnaldo Vieira Moura\thanks{Instituto  de Computacao, Universidade Estadual  de Campinas, 13081\-970  Campinas,  SP.} \and Nadir Matringe \thanks{ Universit\'{e} de Poitiers, Laboratoire Math\'{e}matiques et Applications and Institue de Mathematiques de Jussieu Universit\'{e} Paris 7-Denis Diderot, France.}}

\date{\displaydate{date}}

\maketitle

\begin{abstract}     
We present the notion of \emph{asymptotically non-terminating initial variable values} for linear loop programs. Those  values are directly associated to initial variable values for which the corresponding program does not terminate.  Our theoretical contributions provide us with powerful computational methods for automatically generating sets of asymptotically non-terminating initial variable values.  Such sets are represented symbolically and exactly by a semi-linear space, \emph{e.g.}, characterized by conjunctions and disjunctions of linear equalities and inequalities. Moreover, by taking their complements, we obtain a precise under-approximation of the set of inputs for which the program does terminate.
We can then reduce the termination problem of linear programs to the emptiness check of a specific set of asymptotically non-terminating initial variable values.  Our \emph{static input data analysis} is not restricted only to programs where the variables are interpreted over the reals. We extend our approach and provide new decidability results for the termination problem of affine integer and rational programs.  
\end{abstract}


\section{Introduction}\label{intro}

Proving termination of $\textsf{while}$ loop programs is necessary for the verification of liveness properties that any well behaved and engineered system, or any safety critical embedded system, must guarantee. 
Also, generating input data that demonstrates critical defects and vulnerabilities in programs  allows for new looks at these properties. 
Such crucial \emph{static input data analysis} can be seen as an important trend in the automated verification of loop programs, and is a cornerstone for modern software  industry.
We could list here many verification approaches that are only practical depending on the facility with which termination can 
be automatically determined. 

The \emph{halting problem} is equivalent to the
problem of deciding whether a given program will eventually terminate when
running with a given input. The \emph{termination problem} can be
stated as follows: given an arbitrary program, decide whether the
program eventually halts for every possible input configuration. Both
problems are known to be undecidable \cite{turing}. 
As it happens frequently,
a program may terminate only for a specific set of input data
configurations. The \emph{conditional termination problem} \cite{Cook:2008} asks 
for preconditions representing input data that will cause the program 
to terminate when run with such input data. 

Some recent work on automated termination analysis of imperative loop programs has focused on partial decision procedures based on the discovery and synthesis of ranking functions. 
Such functions map loop variables to well-defined domains where their values decrease further at each iteration of the loop \cite{Colon:2001, Colon02}. 
Several interesting approaches, based on the generation of \emph{linear} ranking functions, have been proposed for loop programs where the guards 
and the instructions can be expressed in a logic supporting linear 
arithmetic~\cite{Bradley05linearranking, Bradley05terminationanalysis}. 
For the generation of such functions, there are effective heuristics \cite{Dams,Colon02} and, in some cases, there are also complete methods for the synthesis of \emph{linear} ranking functions \cite{Podelski04}. 
On the other hand, there are simple linear terminating loop programs for which there is no linear ranking functions. 
Concerning decidability results, the work of Tiwari et al. \cite{tiwaricav04} is often cited when treating linear programs over the reals. 
For linear programs over the rationals and integers, some of those theoretical results have been extended~\cite{Braverman}. 
But the termination problem for \emph{general affine} programs over the integers is left open
in~\cite{Braverman}.  In this article and in our previous work \cite{TR-IC-14-09}, we show that the termination problem for linear/affine program over the integers where the assignments matrix has a real spectrum is decidable. It appears to be the first contribution allowing a constructive and mathematical response to this mentioned open problem. Recently, in \cite{stolen}, the authors were able to address this decidability question for programs with semi-simple and diagonalizable assignments matrices, using strong results from analytic number theory, and diophantine geometry. 
Also, the contributions of this article is not restricted to decidability results, we provide efficient computational methods for new termination and conditional termination analysis.
The framework presented in \cite{Cousot:2012} is devoted to approaches establishing termination by abstract interpretation of termination semantics. The approach exposed in \cite{Cook:2008} searches for non-terminating program executions. They first generates lasso-shaped candidate
paths (i.e., a loop preceded by a finite program path), and then check each path for non-termination. 

The recent literature on \emph{conditional (non-)termination} narrows down to the works presented in
\cite{Gulwani:2008,Cook:2008, Bozga}.
The methods proposed in \cite{Gulwani:2008} allow for the generation of non-linear preconditions.
In \cite{Cook:2008}, the authors derived termination preconditions for simple
programs ---~with only one loop condition~--- by guessing a ranking function
and inferring a supporting assertion. 
Those approaches are sound but not complete.
Also, the interesting approach
provided in \cite{Bozga} (focusing mostly on proofs of decidability), consider several systems and models but is restricted to 
two specific subclasses of linear relations. 
On the one hand, they consider octagonal relations which do not necessarily represent affine loop semantics. 
For such relation there method indicates the use of quantifier elimination techniques and computational steps running in exponential time complexity on the number of variables. 
On the other hand, they treat restricted subclasses of
linear affine relations which must satisfy several restrictions concerning the
associated matrix, such as it being diagonalizable with all
non-zero eigenvalues of multiplicity one.
Using partial termination proofs, the technique proposed in \cite{samir:2013} suggests an incremental proof reasoning on the programs.        
Most directly related work will be discussed in more details (see Sections \ref{practice} and \ref{discus}). 

Despite tremendous progress over the years~\cite{Bradley05Manna, Chen2012, Cousot:2012, Cook:2006, Ben,Gulwani:2008,Cook:2008, 2013:POPL}, the problem of finding a practical, sound and complete method, \emph{i.e.}, an encoding leading deterministically to an algorithm, for determining (conditional) termination remains very challenging.
In this article, we consider linear $\textsf{while}$ loop programs where the loop condition is a conjunction of linear or affine inequalities and the assignments to each  of the variables in the loop instruction block are affine or linear forms. In matrix notation,  \emph{linear or affine loop programs} 
will be represented as: $\textsf{while}\ (F x > b),\ \{x:=Ax+c\}$,
where $A$ and $F$ are matrices, $b$ and $c$ are
vectors over the reals, rationals or integers, and $x$ is a vector of variables  over $\R$, $\Q$, $\N$ 
or $\Z$.  Automated verification for programs presented in a more complex form can often be reduced to the static analysis of a program expressed in this basic affine form. 

We first address  the problem of generating input variable values for which a program does not terminate and, conversely, the problem of obtaining the set of terminating inputs for the same program.  
Initial investigations were reported in~\cite{TR-IC-13-07, TR-IC-13-08} where we discussed  termination analysis algorithms that ran in polynomial time complexity and the initial results on \emph{asymptotically non-terminating initial variable values} ($ANT$, for short) generation where presented in \cite{TR-IC-14-03}. 
Subsequent studies considered the set of $ANT$ whose elements are directly related to input values for which the loop does not terminate~\cite{scss2013rebiha}.
In that work we approached the problem of generating the $ANT$ set for a restricted class of linear programs over the reals, with only one loop condition, and where the associated linear forms of the loop lead to diagonalizable systems with no complex eigenvalues.  
Here, we remove these restrictions.
We show how to handle complex eigenvalues, linear affine programs over $\R$, $\Q$, $\N$ or $\Z$, with conjunctions of several loop conditions, and where the system does not have to be diagonalizable. 
We thus drastically generalize the earlier results in \cite{scss2013rebiha}.  
Further, we introduce new static analysis methods that compute $ANT$ sets
in polynomial time, and also yield a set of initial inputs values for which the program does terminate.  
This attests the innovation of our contributions, \emph{i.e.}, none of the other mentioned works is capable of generating such critical information for non-terminating loops.

We summarize our contributions as follows, with all 
results rigorously stated and proved:\\
\noindent\textbf{Static input data analysis}:\\
\noindent $\bullet$ We introduce the important key
concept of an $ANT$ set. 
Its elements are directly related to initial variables values for which the  program does not terminate.  
Theorems \ref{equivalence}, \ref{reduc-H} and \ref{ant-affine} show the importance of $ANT$  sets. 
 Without loss of generality, we show that the problem of generating $ANT$ sets for the class of affine programs can be reduced to the computation of $ANT$ sets for specific  linear homogeneous programs whose transition matrix has a real spectrum. \\ 
\noindent $\bullet$ We provide efficient computational methods allowing for the exact computation of $ANT$ sets for linear loop programs.  
We automatically generate a set of linear equalities and inequalities describing a semi-linear space that   symbolically and exactly represents such $ANT$ sets.
See Theorem \ref{SUV}. 
 Also, an $ANT$  complement set is a precise under-approximation of the set of terminating inputs for the same program.
Even if these results are mathematical in nature, they are really easy to apply.
 In a practical static analysis scenario, 
one only needs to focus on ready-to-use generic formulas that represent the $ANT$ sets for affine programs. 
See Eqs. (\ref{eqS}), (\ref{eqU}), and (\ref{eqV}).
Such $ANT$ set representations allows for practical computational manipulations --- like union, intersection, and emptiness check ---, and implementations. \\  
\noindent\textbf{Static termination analysis}:\\
\noindent $\bullet$ We obtain  \emph{necessary and sufficient conditions} for the termination of linear programs. 
Further, we reduce the problem of termination for linear programs to the emptiness check of the corresponding $ANT$ set.  
This characterization of terminating linear programs provide us with a deterministic computational procedure to check program termination.  
Such an algorithm is not present in previous works, such as~\cite{tiwaricav04},
that discuss the decidability of the termination problem for linear programs over the reals\\
\noindent\textbf{Decidability results for the termination problem}:\\
\noindent $\bullet$ By extending our results to affine programs over $\Q$, $\N$ and $\Z$, we obtain new decidability results for the program termination problem. 
In \cite{Braverman}, the termination problem for affine programs over the rationals has been proved to be decidable, and the termination of programs over the integers has been proved to be decidable only in the homogeneous case \emph{i.e.}, with loops over the integers and with only one loop condition of the form $\textsf{while}(b x > 0)\{x:=Ax\}$.
Here, we successfully address the question left open in \cite{Braverman}, namely, we settle the decidability problem for program termination in the case of affine programs over the integers under our assumption on the real spectrum $Spec(A)$.
\begin{ex}\label{ex-motivation}\emph{(Motivating Example)} Consider the  program:

\begin{multicols}{2}{
\begin{small}
\begin{lstlisting}
while(x-1/2y-2z>0){
 x:=-20x-9y+75z;
 y:=-7/20x+97/20y+21/4z;
 z:=35/97x+3/97y-40/97z;}
\end{lstlisting}
\end{small}
The initial values of $x$, $y$ and $z$ are represented, respectively, by the parameters $u_1$, $u_2$ and $u_3$. Our prototype outputs the following $ANT$ set:
}
\end{multicols}
\vspace{-10pt}
\noindent\begin{small}
\begin{verbatim}
Locus of ANT:[[u1<-u2+3*u3]]OR[[u1==-u2+3*u3,-u3<u2]]OR
[[u1==4*u3,u2==-u3,0<u3]].
\end{verbatim}  
\end{small}

\noindent $-$ \emph{Static input data analysis}: This semi-linear space  
represents symbolically all asymptotically initial values that are directly associated to initial values for which the program does not terminate. The complement of this set 
 is a precise under-approximation of the set of all initial values for which the program terminates.\\
\noindent $-$ \emph{Termination analysis}: The problem of termination is reduced to the emptiness check of this $ANT$ set. 
 \qed
\end{ex}

Section \ref{prelim} introduces key notations, basic results from linear algebra, and formal programs.
Section \ref{ANT} presents the new notion of \emph{asymptotically non-terminating initial values}, and important results for termination analysis. 
Sections \ref{complex} reduces the study of homogeneous programs to the case where the transition matrix has a real spectrum.
Section \ref{generaltermination} reduces the study of general affine loop programs, with several loop inequality conjunctions,  to that of linear homogeneous programs with one loop condition.
Section \ref{stable} provides new decidability results for the  termination problem of integer and rational affine programs. 
Section \ref{reduced} presents the ready-to-use formulas representing symbolically and exactly the $ANT$ sets for linear homogeneous programs.  
Section \ref{practice} details our computational methods in practice, its algorithm and some experiments. 
We provide a complete discussion in Section \ref{discus}. Finally, 
Section \ref{conclusion} states our conclusions. In the Appendix and in  companion Technical Reports~\cite{TR-IC-14-03, TR-IC-14-09}, we give proofs and details about the computational steps in the running examples.  

\section{Linear Algebra and Linear Loop Programs}\label{prelim}

We recall classical facts from linear algebra.
Let $E$ be a real vector space and let $\A$ belong to $End_\R(E)$, the space of $\R$-linear maps from $E$ to itself.
We denote by $\mathcal{M}(p,q,\R)$ the space of $p\times q$ matrices.
When $p=q$ we may write $\mathcal{M}(p,\R)$. If
$B$ is a basis of $E$, we denote by $A_B=Mat_B(\A)$ the matrix of $\A$ in $B$ in the space $\mathcal{M}(n,\R)$. 
Let $I_n$ be the identity matrix in $\mathcal{M}(n,\R)$, and let $\i$ the identity of $End_\R(E)$. We denote by $det(M)$ the determinant of a matrix in $\mathcal{M}(n,\R)$.
Since $det(A_B)$ is independent of the choice of a basis $B$, we also denote it by $det(\A)$, where $A_B=Mat_B(\A)$. 

\begin{df}\label{char}
The characteristic polynomial of $\A$ is $\chi_\A(T)=det(\A-T \i)$. 
It can be computed as 
$det(A_B-T I_n)$ for any basis $B$ of $E$. Let $Spec_\R(\A)$ be the set of its real roots, which are the real eigenvalues of $\A$. 
If $\l\in Spec_\R(\A)$ has multiplicity $d_\l$ as a root of $\chi_\A$, let $E_\l(\A)=Ker((\A-\l \i)^{d_\l})$ be the generalized eigenspace of $\l$. It contains the eigenspace 
$Ker(\A-\l \i)$, and its dimension is $d_\l$.
\end{df}

We denote by $E^*$ the space of linear maps from $E$ to $\R$.
In the following, we represent linear and affine loop programs in terms of linear forms and their matrix representation. 
\noindent We recall, as it is standard in static program analysis, that a primed symbol $x'$ refers to the next  value of $x$ after a transition is taken. First, we present \emph{transition systems} as representations of imperative programs, and \emph{automata} as their computational models.

\begin{df}
A \emph{transition system} is given by $\langle x, L, \mathcal{T}, l_0,
\Theta \rangle$, where  $x=(x_1, ...,x_n)$ is a set of variables, 
 $L$ is a set of locations and $l_0\in L$ is the initial location. A \emph{state} is given by an interpretation of the variables in $x$. A \emph{transition} $\tau \in \mathcal{T}$ is given by a tuple
  $\langle l_{pre}, l_{post}, q_{\tau}, \rho_{\tau} \rangle$, where $l_{pre}$
  and $l_{post}$ designate the pre- and post- locations of $\tau$, respectively, and
  the transition relation $\rho_{\tau}$ is a first-order assertion over $x
  \cup x'$. The transition guard $q_{\tau}$ is a conjunction of inequalities over $x$.
  $\Theta$ is the initial condition, given as a first-order
  assertion over $x$.
The transition system is said to be \emph{linear} when
$\rho_{\tau}$ is an affine form. 
\end{df}
\noindent We will use the following matrix notations to represent loop programs and their transition systems. We also use simple and efficient procedures to captures the effects of sequential linear assignments into simultaneous updates.    
\begin{df} Let $P=\langle x, l, \mathcal{T}=\langle l, l, q_{\tau}, \rho_{\tau} \rangle, l,\Theta \rangle$, with $x=(x_1,...,x_n)$,
be a loop program. We say that $P$ is a \emph{linear loop program} if:\\
\noindent $\bullet$ Transition guards are conjunctions of linear inequalities. We represent the loop condition in matrix form as $F x > b$ where $F\in\mathcal{M}(m,n,\R)$, and $b\in\R^m$.
By $F x>b$ we mean that each coordinate of  vector $F x$ is  greater than the corresponding coordinate of vector $b$.\\
\noindent $\bullet$ Transition relations 
are affine or linear forms. We represent the linear assignments in matrix form as $x:=Ax+c$, where $A\in\mathcal{M}(n,\R)$, and $c\in\R^n$.
The most general \emph{linear loop program} $P=P(A,F,b,c)$ is defined as $\textsf{while}\ (F x > b),\ \{x:=Ax+c\}$. 
\end{df}    

We will use the following classification. 
\begin{df}\label{classprog}
From the more specific to the more general form:\\
\noindent $\bullet$ \emph{Homogeneous}:  We denote by $P^{\H}$ the set of programs of the form $P(A,f):\textsf{while}\ (f.x > 0), \ \{x:=Ax\}$, where $f$ is a $1\times n$ row matrix corresponding to the loop condition, and $A\in \mathcal{M}(n,\R)$ corresponds to the list of assignments in the loop.\\ 
\noindent $\bullet$ \emph{Generalized Homogeneous}: We denote by $P^{\G}$ the set of programs of the form $P(A,F):\textsf{while}\ (F x > 0), \ \{x:=Ax\}$ where $F$ is a $(m\times n )$-matrix with rows corresponding to the loop conditions ($m$-loop conditions). We will sometimes write $P(A,F)=P(A,f_1,\dots,f_m)$, where the $f_i$'s are the rows of $F$.\\
\noindent $\bullet$ \emph{Affine}:  We denote by $P^{\mathbb{A}}$ the set of programs of the form $P(A,F,b,c):\textsf{while}\ (F x > b), \ \{x:=Ax+c\}$, for $A$ and $F$ as above,  and $b$ and $c\in \R^n$. 
\end{df}
In Section \ref{generaltermination}, we show that the termination analysis for the general class $P^{\mathbb{A}}$ can be reduced to the problem of termination for programs in $P^{\H}$, when transition matrices have a real spectrum.  
\section{The ANT set}\label{ANT}

We present the new notion of \emph{asymptotically non-terminating} ($ANT$) values of a loop program. It will be central in the analysis of non-termination. We start with the definition of the $ANT$ set and then 
give the first important result for homogeneous linear programs. We will extend these results in Section \ref{generaltermination} to generalized linear homogeneous programs
and then expand them further to general affine programs. 
The problem of termination analysis for the general class of linear programs will be reduced to the generation and  the emptiness check of the $ANT$ set for homogeneous linear programs.

Consider the program $P(A,f)$, where
$A\in\mathcal{M}(n,\R)$, $f\in\mathcal{M}(1,n,\R)$. Alternatively, let $\A\in End_\R(E)$, $\f\in E^*$ and the program $P(\A,\f): \textsf{while}\, \f(\x) > 0,  \{\x:=\A \x\}$.
Fixing a basis $B$ of $E$ we will write $A=Mat_B(\A)$, $f=Mat_B(\f)$, $x=Mat_B(\x)$, and so on. We first give the definition of the termination for this class of programs. 

\begin{df}\label{ter}
The program $P(\A,\f)$ terminates on input $\x\in E$ if and only if there exists $k \geq 0$ such that $\f(\A^{k}(\x))$ is not positive. 
Also, 
for $A\in\mathcal{M}_n(\R)$, and $f\in\mathcal{M}_{1,n}(\R)$, we say that $P(A,f)$ terminates on  input $x\in \R^n$ if and only if there exists $k\geq 0$ such that $f A^{k} x$ is not positive. Thus,  $P(\A,\f)$ is non-terminating if and only if there
exists an input $\x\in E$ such that $\f(\A^{k}(\x)) > 0$ for all
$k\geq 0$. In matrix terms, $P(A,f)$ is non-terminating on input
 $x\in \R^n$ if and only if $\langle A^k x, f \rangle > 0$ for
all $k\geq 0$.\qed
\end{df}
 
The following lemma is obvious.

\begin{LM}
$P(\A,\f)$ terminates on $\x$ if and only if $P(A,f)$ terminates on $x$.\qed
\end{LM}

Next, we introduce the important notion of an asymptotically non-terminating value. 

\begin{df}\label{ant}
We say that $\x\in E$ is an asymptotically non-terminating value for 
$P(\A,\f)$ if there exists $k_\x\geq 0$ such that $P(\A,\f)$ is non-terminating on 
$\A^{k_\x}(\x)$. We will also say that $\x$ is ANT for $P(\A,\f)$. We will also say that $P(\A,\f)$ is ANT on $\x$. \qed
\end{df}
The definition of 
an ANT value for a program $P(A,f)$ with $A\in\mathcal{M}_n(\R)$ and $f\in\mathcal{M}_{1,n}(\R)$ is similar. It is again 
obvious that $\x$ is ANT for $P(\A,\f)$ if and only if $x$ is ANT for $P(A,f)$. 
If $K$ is a subset of $E$, we will say that 
$P(\A,\f)$ is ANT on $K$ if it is ANT on every $\x$ in $K$.
\noindent Note that $P(\A,\f)$ is non-terminating on $\A^{k_\x}(\x)$ if and only if $\f(\A^k(\x))$ is $>0$ for $k\geq k_\x$.
The following example illustrates $ANT$ sets and their properties.

\begin{ex} Consider again Example\ref{ex-motivation}.\\ 
It is easy to check that the initial value $u=(-9,3,-2)^{\top}$ belongs to the $ANT$ set. But the program terminates on $u$ because with this initial value no loop iteration will be performed as $fA^0 u=-13/2$. From Definition \ref{ant}, there exists $k_{u}\geq 0$ such that $P(A,f)$ is non-terminating on $A^{k_u}u$. Also, $fA^1u=-5/2$, but $fA^2u>0$ and the program is non-terminating on $A^2u = (63,3,22)^{\top}$. So, the input $u=(-9,3,-2)^{\top}$ is $ANT$ for $P(A,f)$ with $k_u=2$.
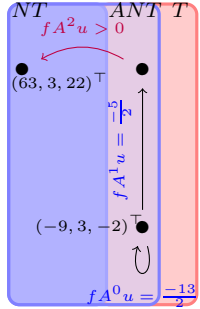
\begin{tikzpicture}

\tikzstyle{sectionT}=[draw, rectangle, rounded corners, very thick, minimum height=4cm, minimum width=1.3cm, draw=red!50,fill=red!20, anchor=south west]
\tikzstyle{section}=[draw, rectangle, rounded corners, very thick, minimum height=4cm, draw=blue!50, minimum width=2cm, fill=blue!20, anchor=south west, fill opacity=0.5]
\tikzstyle{sectionNT}=[draw, rectangle, rounded corners, very thick, minimum height=4cm, minimum width=1.3cm,  draw=blue!50, fill=blue!40, anchor=south west]

\node[sectionT] (a) at (-8.8, 0) {};
\node [sectionNT](c) at (-10, 0) {};
\node [section] (b)  at (-10, 0) {};



\node [above] (u0) at (-8.2,0.8) {$\bullet$};
\node [above] (u1) at (-8.2,2.9) {$\bullet$};
\node [above] (u2) at (-9.8,2.9) {$\bullet$};

\node [above] (nt) at (-9.7,3.7) {\begin{scriptsize}$NT$\end{scriptsize}};
\node [above] (ant) at (-8.3,3.7) {\begin{scriptsize}$ANT$\end{scriptsize}};
\node [above] (t) at (-7.7,3.7) {\begin{scriptsize}$T$\end{scriptsize}};
\node [above] (um) at (-8.9,0.8) {\begin{tiny}$(-9,3,-2)^{\top}$\end{tiny}};
\node [above] (um) at (-9.3,2.7) {\begin{tiny}$(63,3,22)^{\top}$\end{tiny}};

\draw[->] (u0) to node[above,sloped,blue] {\begin{tiny}$fA^1u=\frac{-5}{2}$\end{tiny}} (u1);
\draw[->] (u0) to [loop below] node[below,blue] {\begin{tiny}$fA^0u=\frac{-13}{2}$\end{tiny}} (u0);
\draw[->,purple] (u1) to [bend right] node[above,sloped] {\begin{tiny}$fA^2u>0$\end{tiny}} (u2);

\node [left, text width=10cm,align=justify] at (-2,2) {
};
\end{tikzpicture}

\qed
\end{ex}

We denote by $ANT(P(\A,\f))$ the set of $ANT$ values of $P(\A,\f)$, and similarly for programs involving matrices. 
From now on, we give definitions and statements in terms of 
programs involving linear maps, and let the reader infer the obvious adaptation for programs involving matrices. If the set $ANT(P(\A,\f))$ is not empty, we say that the program $P(\A,\f)$ is $ANT$. We will also write $NT$ for non terminating. The following theorem already shows the importance of $ANT$ sets:  termination for linear programs is reduced to the emptiness check of the $ANT$ set. 

\begin{thm}\label{equivalence}
The program $P(\A,\f)$ in $P^{\H}$ is $NT$ if and only if it is $ANT$ (i.e., $ANT(P(\A,\f))\neq \emptyset $). More generally, if $K$ is an $\A$-stable 
subset of $E$, the program $P(\A,\f)$ is $NT$ on $K$ if and only if it is $ANT$ on $K$.\qed
\end{thm}  
 
We just saw that the set of $NT$ values is included in the  $ANT$ set, but the most important property of an $ANT$ set resides in the fact that each of its elements gives an associated element in $NT$ for the corresponding program. That is, each element $\x$ in the $ANT$ set, even if it does not necessarily belong to the $NT$ set, refers directly to initial values $\A^{k_\x}(\x)$ for which the program does not terminate. Hence there exists a number of loop iterations $k_\x$, departing from the initial value $\x$, such that $P(\A, \f)$ does not terminate on $\A^{k_\x}(\x)$. This does not imply that $\x$ is $NT$ for $P(A,f)$ because the program $P(A,f)$ could terminate on $\x$ by performing a number of loop iterations strictly smaller than $k_\x$. 
On the other hand, the $ANT$ set is more than an over-approximation of the $NT$ set, as it will provide us with a deterministic and efficient way to decide termination. 

Let ${ANT}^c$ be the complement of the $ANT$ set. It gives us an under approximation for the set of all initial values for which the program terminates. 

\begin{cor}\label{co-ANT}
Let $P(\A,\f)$ be  in $P^{\H}$.
Then $P(\A,\f)$ terminates on the complementary set $ANT^c(P(\A,\f))$ of $ANT(P(\A,\f))$.   \qed
\end{cor}
 
\section{ANT set and Complex Eigenvalues}\label{complex}  

Let $Spec(\A)$ be the set of all eigenvalues  of $\A$. We show that the non-real eigenvalues in $Spec(\A)$ do not affect the static termination analysis for a linear homogeneous program. 
We also show that the problem of checking the termination of linear programs can be reduced to verifying whether $Spec_\R(\A)\subset Spec(\A)$. 
This will provide a complete and deterministic procedure to 
statically and automatically verify  the termination of a linear loop program. 
We will  reduce the problem of termination for a linear program $P(\A,\f)$ to the emptiness check of $ANT^r(P(\A,\f))$, the set of 
$ANT$ values in the maximal $\A$-stable subspace $E^r$ such that the restriction $\A_{|E^r}$ has only real eigenvalues, that is,
such that $Spec(\A_{|E^r})= Spec_\R(\A)$.
We will also show how to compute the $ANT^r(P(\A,\f))$ sets.
The complement of $ANT^r$ will turn out to be the set of initial values for which the program does terminate.  

In $\R[X]$ the characteristic polynomial $\chi_{\A}$
 factors uniquely as
$\chi_{\A}^{nr}\prod_{\l \in Spec_\R(\A)}(X-\l)^{d_\l}$, where $\chi_{\A}^{nr}$ has no real roots. We denote by $\chi_{\A}^+$ the product $\prod_{\l>0 \in Spec_\R(\A)} (X-\l)^{d_\l}$, and by $\chi_{\A}^-$
 the product $\prod_{\l<0 \in Spec_\R(\A)} (X-\l)^{d_\l}$. We recall that for $P\in \R[X]$ the spaces  $Ker(P(A))$ and $Im(P(A))$ are always $A$-stable.

\begin{df}
We denote by $E^+$ the space $Ker(\chi_{\A}^+(\A))$, by $E^-$ the space $Ker(\chi_{\A}^-(\A))$, by $E^r$ the space 
$E^+ \oplus E_0(\A) \oplus E^-$, and by $E^{nr}$ the space $Ker(\chi_{\A}^{nr}(\A))$. They are all $\A$-stable.\qed
\end{df}

The space $E^r$ is such that $\A_{|E^r}$ has only real eigenvalues, that is, $Spec(\A_{|E^r})=Spec_\R(\A)$. We recall the following proposition from basic linear algebra, 
which is a consequence of the fact that if the gcd $P\wedge Q$ of two polynomials in $\R[T]$ is equal to $1$, then $Ker(PQ(A))=Ker(P(A))\oplus Ker(Q(A))$.

\begin{prop}
One has the decompositions:\\ 
$E^+=\oplus_{\l>0\in Spec(\A)} E_{\l}(\A)$, $E^-=\oplus_{\l<0 \in Spec(\A)} E_{\l}(\A),$ and $E=E^r\oplus E^{nr}$.\qed
\end{prop}

We also recall a theorem from \cite{TR-IC-14-08} (see Theorem 3.3 and 3.4 from \cite{TR-IC-14-08}), which gives a necessary and sufficient condition for $P(\A,\f)$ to be terminating. In \cite{TR-IC-14-08}, the result is rigorously stated in solid mathematical way (i.e, a mix of topological and algebraic arguments).
\begin{thm}\label{nadthm}
The program $P(\A,\f)$ is non-terminating if and only if there is a $\l>0$ in $Spec(\A)$, such that $E_\l(\A)\not\subset Ker(\f)$.\qed
\end{thm}
Theorem \ref{nadthm} has the following important consequence.
\begin{prop}\label{x+}
If  program $P(\A,\f)$ is $ANT$ on $\x$ in $E$, where $\x=\x^+ +\x'$ with  $\x^+\in E^+$ and 
$\x'\in E'=E^-\oplus E_0(\A)\oplus E^{nr}$, then it is asymptotically non-terminating on $\x^+$.\qed
\end{prop}

We can refine  Proposition \ref{x+} using the following result.

\begin{thm}\label{xr}
If  program $P(\A,\f)$ is asymptotically non-terminating on $\x=\x^r +\x^{nr}$, where $\x^r\in E^r$ and 
$\x^{nr}\in E^{nr}$, then it is asymptotically non-terminating on $\x^{r}$.\qed
\end{thm}

We can now state the next result.

\begin{thm}\label{thmcomplex}
We write $\A^r$ for the restriction of $\A$ to $E^r$, $\f^r$ for the restriction of $\f$ to $E^r$, and let $ANT^r(P(\A,\f))=ANT(P(\A,\f))\cap E^r$.
 Then $ANT(P(\A,\f))$ is non empty if and only if $ANT^r(P(\A,\f))$ is non empty. In particular,  $P(\A,\f)$ terminates if 
 and only if  $P(\A^r,\f^r)$ terminates. In any case, one has $ANT^r(P(\A,\f))=ANT(P(\A^r,\f^r))$.\qed 
\end{thm}

\begin{ex}
Consider the following program and the associated matrices:
\vspace{-5pt}
\begin{multicols}{2}{
\begin{small}
\begin{lstlisting}
while(3t+7s+x-1/2y-2z>0){
 t:=t-s;
 s:=t+2s; 
 x:=-20x-9y+75z;
 y:=-7/20x+97/20y+21/4z;
 z:=35/97x+3/97y-40/97z;}
\end{lstlisting}

\[
A'=\left(\begin{array}{*5{c}}

\tikzmark{1}{$1$}  &  -1              &                 &        &  \\
             1   &   \tikzmark{2}{$1$}&                 &        &  \\
                 &                  & \tikzmark{3}{}  &        &  \\
                 &                  &                 &    A    &  \\
                 &                  &                 &       & \tikzmark{4}{}  
\end{array}\right)
\]
  \Highlight{1}{2}{blue} 
  \Highlight{3}{4}{red} 
and $f'=(3,7, f)^{\top}$. Here the submatrices $A$ and $f$ are the one associated to Example \ref{ex-motivation}. 
\end{small}
}
\end{multicols}
\vspace{-10pt}
\noindent 
The submatrix $A=\begin{small}\begin{pmatrix}-20&-9&75\\7&8&-21\\-7&-3&26\end{pmatrix}\end{small}$ correspond to the simultaneous updates representing the the sequential loop assignments of Example \ref{ex-motivation}.
The submatrix $\begin{small}\begin{pmatrix} 1& -1\\ 1& 1\end{pmatrix}\end{small}$ encodes the effect of the two sequential instructions $\textsf{t:=t-s;}$ and $\textsf{t:=t+2s;}$ in terms of simultaneous updates. As this submatrix has only non real complex eigenvalues, Theorem \ref{thmcomplex} says that $ANT^r(P(A',f'))=ANT(P(A,f))$. 
If the initial values of $t,s,x,y$ and $z$ are represented, respectively, by the parameters $u_1,u_2,u_3,u_4$ and $u_5$, then we obtain conditions similar to that in Example \ref{ex-motivation}. But now the $ANT$ locus is described by the parameters $u_3,u_4$ and $u_5$, thus: \begin{small}
\begin{verbatim}
[[u3 < -u4 + 3*u5]] OR [[u3 == -u4 + 3*u5, -u5 < u4]] OR 
[[u3 == 4*u5, u4 == -u5, 0 < u5]]
\end{verbatim}
\end{small}\qed
\end{ex}

\section{The ANT set for affine programs}\label{generaltermination}

In the first subsection we express the $ANT$ set of programs $P(\A,\f_1,\dots,\f_m)\in P^{\G}$, involving several linear forms $\f_i$, in terms 
of the sets $ANT(P(\A,\f_i))$. 
In the next subsection we reduce the computation of the $ANT$ sets and  termination  of affine programs $P(\A,(\f_i)_{i=1,\dots,m},\b,\c)\in P^{\mathbb{A}}$ to the 
corresponding problems for programs in $P^{\G}$.
Hence, after  Subsection \ref{generalized}, to corresponding problems 
for programs $P(\A,\f)$ where $\A$ has a real spectrum.   

\subsection{Handling generalized homogeneous programs}\label{generalized}

Consider $P(\A,\F)=P(\A,(\f_i)_{i=1,\dots,m})$ in $P^{\G}$ in the form $\textsf{while}\ (\forall i=1,\dots,m,\ \f_i \x > 0), \ \{\x:=\A\x\}$. 
We start with the following lemma. 

\begin{LM}
The value $\x$ is $NT$ for $P(\A,\F)$ in $P^{\G}$ if and only if it is $NT$ for all $P(\A,\f_i)$ with $i\in \{1,\dots,m\}$.\qed 
\end{LM}

Now, we define $ANT$ values  for such programs.

\begin{df}\label{antG}
We say that $\x$ is $ANT$ for $P(\A,\F)$ if there exists $k_\x$ such that for all $i\in\{1,\dots , m\}$ we have $\f_i(\A^k(\x)) > 0$ for 
$k>k_\x$, that is, if $\x$ is $ANT$ for all programs $P(\A,\f_i)$.\qed
\end{df}  

Again we have the following easy but important lemma.

\begin{LM}\label{ANTandNT}
Program $P(\A,\F)$ is $NT$ if and only if it is $ANT$, that is, $ANT(P(\A,\F))\neq \emptyset$.\qed
\end{LM}

The next result will be crucial for the main result of this section.

\begin{prop}\label{crux}
If  program $P(\A,\F)$ is $ANT$ then we must have  
$\underset{i}{\bigcap} ANT^r(P(\A,\f_i))\neq \emptyset$.\qed
\end{prop}

For the main result of this section, let $ANT^r(P(\A,\F))=\underset{i}{\bigcap} ANT^r(P(A,f_i))$.

\begin{thm}\label{reduc-H}
Program $P(\A,\F)$ is non-terminating if and only if $ANT^r(P(\A,\F))\neq \emptyset$. 
In particular, this gives a deterministic procedure to check if $P(\A,\F)$ is NT, as we can always compute 
$ANT^r(P(\A,\F))$. Moreover, if $P(\A,\F)$ is non-terminating we obtain an under approximation of the set 
of terminating values for $P(\A,\F)$, namely $E^r-ANT^r(P(\A,\F))$.\qed
\end{thm}

\subsection{Generalization to affine programs}\label{antaffine}

We now reduce the affine case to the homogeneous case. First we define the notion of $ANT$ values for this class of programs.  
\begin{df}
Let $P(A,F,b,c)$ be an affine program in
$P^{\mathbb{A}}$. For $x=x_0\in \R^n$, denote by $x_1$ the vector $Ax+ c$, and recursively let  
$x_k=Ax_{k-1}+c$. We say that a vector $x$ is $ANT$ for $P(A,F,b,c)$ if there is some $k_x$ such that $k\geq k_x$ implies $Fx_k>b$. 
We denote by $ANT(P(A,F,b,c))$ the set of $ANT$ inputs of $P(A,F,b,c)$, and we set $ANT^r(P(A,F,b,c))=ANT(P(A,F,b,c))\cap E^r$.\qed 
\end{df}

Let $A\in \mathcal{M}(n,\R)$, $F\in \mathcal{M}(m,n,\R)$, $b= (b_1,\dots , b_m)^{\top}$  in $\mathcal{M}(1,m,\R)$, and $c$ a
vector in $\mathcal{M}(1,n,\R)$. We denote by $P(A,F,b,c)\in
P^{\mathbb{A}}$ the program which computes $x:=Ax+c$ as long as $Fx>b$.
Now we define $A'\in \mathcal{M}(n+1,\R)$ and $V'\in
\mathcal{M}(m+1,n+1,\R)$ as follows:
$A'=\begin{bmatrix} A &c \\ 0 & 1\end{bmatrix}$,
$F'=\begin{bmatrix} F &-b \\ 0 & 1\end{bmatrix}$. The following theorem shows that the generation of the $ANT$ set for a program in $P^{\mathbb{A}}$ reduces to the generation of the $ANT$ set for an associated program in $P^{\H}$.      

\begin{thm}\label{ant-affine}
Let $P(A,F,b,c)$ be an affine program in $P^{\mathbb{A}}$. Consider the matrices $A'$ and $F'$ as constructed just above. A vector $x$ is $ANT$ for $P(A,F,b,c)$ if and only if the vector $\begin{bmatrix} x \\ 1 \end{bmatrix}$ is $ANT$ for $P(A',F')$.\qed
\end{thm}
\section{New Decidability Results for Affine Rationals and Integers Programs}\label{stable}

In this section, we provide new decidability results for the termination of affine programs over the rationals and the integers when the transition matrix has a real spectrum. This problem has been studied in \cite{Braverman}, where the termination of programs over
the rationals has been proved to be decidable in general, and over the integers
it has been proved to be decidable only in the
homogeneous case. 
With our restriction on the spectrum, 
 we obtain termination decidability for general affine programs $P(A,F,b,c)$ over the
rationals and the integers.
Moreover, this comes
with an exact algorithm to check termination. Such a simple and clear
algorithm is not provided in~\cite{Braverman}.
In addition, we settle an open question in~\cite{Braverman}, 
namely the decidability of termination in the case of
affine programs over the integers --- under our assumption on $Spec(A)$. 
Further,
our method works for a very large family
of subsets of $\R^n$ --- and not only of $\Q^n$, $\Z^n$, or $\N^n$, --- namely, it works for  all those  stable subsets
under $x\mapsto Ax+c$.  Formally, we show that the computation of the $ANT$ set of an affine program $P(A,F,b,c)$ gives a very simple answer to the termination 
problem on any subset $K$ of $\R^n$ which is stable under $x\mapsto Ax+c$. However, in the previous sections, 
we only determine the set $ANT^r(P(A,F,b,c))\subset ANT(P(A,F,b,c))$. While this is sufficient to answer the termination problem on the reals in general, it is not enough to answer the termination on stable subspaces like $K$, as in general, 
we do not have $K=K\cap E^r\oplus K\cap E^{nr}$. 

For now, we will need to restrict ourselves to the case where $Spec(A)$ is real. 
We will remove this restriction and raise the general problem
in another companion article where more technical details will
be presented together with some experiments.\\

\begin{df}
If $P(A,F,b,c)$ is an affine program.
Let  $A'$ and $F'$ be the matrices defined in the previous section. Then $ANT^r(P(A,F,b,c))$ is the set of elements 
$x\in \R^n$ such that $x'=\begin{bmatrix} x \\ 1 \end{bmatrix} \in ANT^r(P(A',F'))$.\qed
\end{df}

\begin{prop}\label{real2real} If  $A$ has a real spectrum, then we must have $ANT(P(A,F,b,c))=ANT^r(P(A,F,b,c))$.\qed \end{prop}

Here is the core theoretical contribution of this section.

\begin{thm}\label{K}
Let $K$ be a subspace of $\R^n$, stable under $x\mapsto Ax+c$. Then $P(A,F,b,c)$ is non-terminating on $K$ if and only if 
$ANT(P(A,F,b,c))\cap K\neq \emptyset$. 
Moreover, the program terminates on $K-ANT(P(A,F,b,c))\cap K$.\qed
\end{thm}

When $Spec(A)$ is real we must have $ANT^r(P(A,F,b,c))=ANT(P(A,F,b,c))$ by Proposition \ref{real2real}.
Hence, we obtain the following corollary.

\begin{cor}
When $Spec(A)$ is real the termination problem on $K$ is decidable.
Additionally we can always compute the set $ANT(P(A,F,b,c))\cap K$.\qed 
\end{cor}


In~\cite{Braverman}, the termination of affine programs over the integers was left open. See Section \ref{discus} for a discussion. Our criterion for termination over stable subspaces allows us to answer this question when $A$ has a real spectrum.  We have the following new characterization decidability for termination of affine programs  over the rationals and the integers.

\begin{cor}
If $A$ and $c$ have rational, respectively integer, coefficients, and $Spec(A)$ is real, then $P(A,F,b,c)$ terminates 
over $\Q^n$, respectively $\Z^n$, if and only if $ANT(P(A,F,b,c))\cap \Q^n=\emptyset$, respectively $ANT(P(A,F,b,c))\cap \Z^n=\emptyset$.\qed
\end{cor}

\section{Automatic generation of ANT values}\label{reduced}
We show how to compute exactly $ANT$ loci. As we saw in the previous sections, it is enough to treat the 
case where the program belongs to $P^{\H}$, \emph{i.e.}, can be written as $P(\A,\f)$, with $Spec(\A)$ a set of reals. 
In Subsection \ref{regular} we start with what we call the regular case:  a condition on $\A$ and $\f$ which is satisfied most of the time.
In Subsection \ref{real} we explain how the general case 
reduces to the regular case. Given that we can 
generate the $ANT$ locus of any $P(\A,\f)$ in $P^{\H}$ when $Spec(\A)$ is a set of reals, we are in fact able to generate $ANT^r(P(A,F,b,c))$ for any $P(A,F,b,c)$ in $P^{\mathbb{A}}$, without any further hypothesis about $Spec(A)$. In this section, we drop the bold font when denoting elements in $E$.

\subsection{The regular case}\label{regular}
In this subsection we assume that $K(\A,\f)=\cap_{k\geq 0} Ker(\f\circ \A^k)=\cap_{k= 0}^{n-1} Ker(\f\circ \A^k)$, and $E_0(\A)$ are simply $\{0\}$. We first recall the following consequence of the existence of a Jordan form for $\A$. 

\begin{LM}\label{goodbasis}
Let $\l$ be a nonzero eigenvalue of $\A$. We can produce a basis $B_\l$ of $E_\l(\A)$ such that $Mat_{B_\l}(\A)$ is of the form 
$\l. diag(T_{\l,1},\dots,T_{\l,r_\l})$, where each $T_{\l,i}$ is a matrix of size $n_{\l,i}$  of the form 
\begin{scriptsize}$\begin{pmatrix} 1 & 1 &  &   &   &  \\
  & 1 & 1&   &   &  \\
  &  & \ddots &\ddots &   & \\
  &  &  & 1 & 1  & \\
  &  &  &  & 1  & 1 \\
    &  &  &  &   & 1
\end{pmatrix}$\end{scriptsize}, and with $n_{\l,i}\leq n_{\l,i+1}$ for $i$ between $1$ and $r_\l-1$. 
\end{LM}

In fact, we assumed $K(\A,\f)$ to be null, there is only one block.

\begin{prop}\label{oneblock}
For every $\l$ in $Spec(\A)$ with $\l\neq 0$, we have  $Mat_{B_\l}(\A_{|E_{\l}(A)})=\l.T_{\l,1}$.
So, we simply write 
$T_\l=T_{\l,1}$. 
\end{prop}

Now we compute the power of $T_\l$.

\begin{LM}\label{poweroflambda}
$(T_{\l})^k= \left(\begin{smallmatrix} 1 & (_1^{k}) & (_2^{k})  & \dots   & \dots   & (_{d_{\l}-1}^{k}) \\
  & 1 & (_1^{k})& (_2^{k})  & \ddots   & \vdots \\
  &  & \ddots &\ddots &  \ddots & \vdots \\
  &  &  & 1 & (_1^{k})  & (_2^{k})\\
  &  &  &  & 1  & (_1^{k}) \\
      &  &  &  &   & 1
\end{smallmatrix}\right).$
\end{LM}

 We write $B=B_{\l_1} \cup \dots \cup B_{\l_t}$.
 It is a basis of $E$. 
We write $Mat_B(f)$ as $(a_{\l_1,1},a_{\l_1,2},\dots,a_{\l_1,d_{\l_1}},\dots,a_{\l_t,1},\dots,a_{\l_t,d_{\l_t}})$. 
The previous lemma has the following consequence.

\begin{prop}\label{version1}
For $\l$ in $Spec(\A)$, there are well determined polynomials $P_{\l,j}\in \R[X]$, for
$j$ between $1$ and $d_{\l}$, such that $Mat_B(\f\circ \A^k)=(P_{\l,1}(k),\dots,P_{\l,d_{\l}}(k))$. In fact, 
$P_{\l,j}(k)=a_{\l,1} (_{j-1}^{k})+a_{\l,2} (_{j-2}^{k})+\dots+a_{\l,j}$.
In particular, $P_{\l,j}$, as a polynomial in $k$, is of degree at most $j-1$. We thus write it as 
$P_{\l,j}(k)=b_{\l,j-1}^j k^{j-1}+\dots+b_{\l,1}^j k+b_{\l,0}^j$,
where we can compute each $b_{\l,l}^j$ explicitly as a linear combination of the $a_{\l,i}$'s.
\end{prop}

We now give a procedure, in several steps, to determine the set of $ANT$ values. 
Here, in order to lighten the notations, for $x=\sum_{\l\in Spec(\A)} x_\l$, we write $P_\l(x_\l,k)=\sum_{j=1}^{d_\l} x_{\l,j} P_{\l,j}(k)$. By convention, if $\l$ is not an eigenvalue, the polynomial $P_{\l,j}$, so that $P_\l(x_\l):k\mapsto P_\l(x_\l,k)$, is zero. We set $b_{\l,i}^{j}=0$ as soon as $i> d_\l$. We obtain the expression 
$P_\l(x_\l,k)=\sum_{j=0}^{d_\l-1} \phi_{\l,j}(x_\l)k^{j}$,
where 
\begin{align*}
\phi_{\l,j}(x_\l)=& b_{\l,j}^{j+1}x_{\l,j+1}+b_{\l,j}^{j+2}x_{\l,j+2}x_{\l,j+2}+\dots\\
&+b_{\l,j}^{d_\l}x_{\l,d_\l}.
\end{align*} 
We set $\phi_{\l,j}(x_\l)=0$ as soon as $j\geq d_\l$.  We also write $Q_{\pm\l }^+(x_{\pm\l})=P_{|\l|}(x_{|\l|})+P_{-|\l|}(x_{-|\l|})$, and $Q_{\pm\l}^-(x_{\pm\l})=P_{|\l|}(x_{|\l|})-P_{-|\l|}(x_{-|\l|})$.
In particular, we have 
$$Q_{\pm\l }^+(x_{\pm\l},k)=\sum_{j=0}^{e_{|\l|}-1} \phi_{\pm\l,j}^+(x_{|\l|},x_{-|\l|})k^{j},$$
 where $e_{|\l|}=max(d_{|\l|},d_{-|\l|})$, 
\begin{equation*}\label{phi+}
\phi_{\pm\l,j}^+(x_{|\l|},x_{-|\l|})=\phi_{|\l|,j}(x_{|\l|})+\phi_{-|\l|,j}(x_{-|\l|}),
\end{equation*} 
and
\begin{equation*}\label{Q-}
Q_{\pm\l}^-(x_{\pm\l},k)=\sum_{j=0}^{e_{|\l|}-1} \phi_{\pm \l,j}^-(x_{|\l|},x_{-|\l|})k^{j},
\end{equation*}
where 
$$\phi_{\pm\l,j}^-(x_{|\l|},x_{-|\l|})=\phi_{|\l|,j}(x_{|\l|})-\phi_{-|\l|,j}(x_{-|\l|}).$$

As the first step of the procedure, we give a set of constraints giving birth to $ANT$ values.

\begin{prop}\label{1}
Let $\l$ be a positive eigenvalue, such that both $Q_{\pm\l}^+(x_{|\l|})$ and $Q_{\pm\l}^-(x_{|\l|})$  have a positive dominant term, and such that 
$P_{\mu}(x_\mu)$ is zero whenever $|\mu|>\l$. Then $x$ is an ANT point of $P(\A,\f)$.
\end{prop}

In terms of the linear forms $\phi_{|\l|,j}$ and $\phi_{|\l|,j}^\pm$ we have the following proposition.

\begin{prop}\label{1'}
For $\l>0$ in $Spec(\A)$, and two integers $k$ and $k'$ between $0$ and $d_\l-1$,  denote 
by $S_{k,k'}^{\l}$ the set of $x$ in $E$ which satisfy:\\
1) for all $\mu$ with $|\mu|>\l$, and all $j$ between $0$ and $d_\mu-1$: $\phi_{\mu,j}(x_\mu)=0$.\\
2) for all $e_\l-1\geq j>k$: $\phi_{\pm\l,j}^+(x_\l,x_{-\l})=0$.\\
3) for all $e_\l-1\geq j>k'$: $\phi_{\pm\l,j}^-(x_\l,x_{-\l})=0$.\\
4) $\phi_{\pm\l,k}^+(x_\l,x_{-\l})>0$.\\
5) $\phi_{\pm\l,k'}^-(x_\l,x_{-\l})>0$.\\
Consider the set $\Delta_S=\{(\l, k, k')| \l>0 \in Spec(\A), k\in \{1,\dots,d_\l-1 \},k'\in \{1,\dots,d_\l-1 \}\}$. If $x$ belongs to 
\begin{equation}\label{eqS}S=\bigvee_{(\l, k, k')\in\Delta_S} S_{k,k'}^\l,\end{equation} 
then $x$ is $ANT$.
\end{prop}

We illustrate how to generate the constraints of Proposition \ref{1'} on a running example.

 \begin{ex}\label{exS} \emph{(Running example)}
Let $T=Mat_B(\A)$, and $f=Mat_B(\f)$ be as follows:
 $\begin{scriptsize}T=\begin{pmatrix} 
 1 & 1 & 0 & 0 & 0 & 0 \\
0 & 1 & 0 & 0 & 0 & 0 \\
 0 & 0 & -1 & -1 & 0 & 0 \\
 0 & 0 & 0 & -1 & 0 & 0 \\
 0 & 0 & 0 & 0 & 2 & 0 \\
 0 & 0 & 0 & 0 & 0 & -2                
\end{pmatrix},\end{scriptsize}$ and  $\begin{scriptsize}f=\begin{pmatrix}1\\1\\1\\1\\1\\1\end{pmatrix}\end{scriptsize}$. Also, $f=(a_{1,1},a_{1,2}, a_{-1,1}, a_{-1,2}, a_{2,1}, a_{-2,1})$.

In~\cite{TR-IC-14-03} we give values of all the needed terms 
appearing in the computation of the $ANT$ set. We have four eigenvalues: $1$ of multiplicity $2$, $-1$ of multiplicity $2$, $2$ of multiplicity $1$ and $-2$ of multiplicity $1$. 
For Proposition \ref{1'}, we take only the positive eigenvalues $2$ and $1$. From the multiplicity of these eigenvalues, we know that we have to consider $k\in\{0,1\}$ and $k'\in\{0,1\}$. 
From Proposition \ref{1'} we generate: 
\begin{small}
\begin{multline*}
S^2_{0,0} \equiv (x_{2,1}+x_{-2,1}>0) \wedge (x_{2,1}-x_{-2,1}>0)
\end{multline*}
\vspace*{-5ex}
\begin{multline*}
 S^1_{0,0}\equiv(x_{2,1}=0)\wedge (x_{-2,1}=0) \wedge  (x_{1,2}+x_{-1,2}=0)\\ \wedge (x_{1,2}-x_{-1,2}=0)\wedge (x_{1,1}+x_{1,2}+x_{-1,1}+x_{-1,2}>0)\\
 \wedge (x_{1,1}+x_{1,2}-x_{-1,1}-x_{-1,2}>0)
\end{multline*} 
\vspace*{-5ex}
\begin{multline*}
S^1_{0,1}\equiv   (x_{2,1}=0) \wedge (x_{-2,1}=0) \wedge (x_{1,2}+x_{-1,2}=0) \\
  \wedge (x_{1,1}+x_{1,2}+x_{-1,1}+x_{-1,2}>0)
   \wedge (x_{1,2}-x_{-1,2}>0)
\end{multline*}
\vspace*{-5ex}
\begin{multline*}
S_{1,0}^1\equiv  (x_{2,1}=0) \wedge (x_{-2,1}=0) \wedge    (x_{1,2}+x_{-1,2}>0)\\ \wedge  (x_{1,1}+x_{1,2}-x_{-1,1}-x_{-1,2}>0).
\end{multline*}
\end{small}  
  
Proposition \ref{1'} gives the  set $S= S^2_{0,0} \vee S^1_{0,0} \vee S^1_{0,1} \vee S_{1,0}^1 \vee S_{1,1}^1$.
 \end{ex}
 
The next result gives other set of constraints for $ANT$ values. 

\begin{prop}\label{2}
Let $\l$ be a positive eigenvalue and $|\l'|<\l$ in $|Spec(\A)|$ such that:\\
\noindent $\bullet$ $P_{\mu}(x_\mu)$ is zero whenever $|\mu|>\l$,\\
\noindent $\bullet$ $Q_{\pm\l}^-(x_{\pm\l})=0$,\\
\noindent $\bullet$ $Q_{\pm\l}^+(x_{\pm\l})=2P_\l(x_\l)$ 
has a positive dominant term,\\
\noindent $\bullet$ $Q_{\pm\mu}^-(x_{\pm\mu})$ is zero whenever $|\l'|<|\mu|<\l$,\\
\noindent $\bullet$ $Q_{\pm\l'}^-(x_{\pm\l'})$ has a positive dominant term\\
Then $P(\A,f)$ is ANT on $x$.
\end{prop}

In terms of the linear forms $\phi_{\l,j}$, $\phi_{\pm\l,j}^+$ and  $\phi_{\pm\l,j}^-$:

\begin{prop}\label{2'}
For $\l>0$ and $\l'$ in $Spec(\A)$, with $|\l'|<\l$, an integer $k$ between $1$ and $d_\l-1$, and an integer 
$k'$ between $1$ and $d_{\l'}-1$, we denote 
by $U_{k,k'}^{\l,|\l'|}$ the set of $x$ in $E$ which satisfy:\\
1) for all $\mu$ with $|\mu|>\l$, and all $j$ between $0$ and $d_\mu-1$: $\phi_{\mu,j}(x_\mu)=0$.\\
2) for all $0\leq j \leq e_\l-1 $: $\phi_{\pm\l,j}^-(x_{|\l|},x_{-|\l|})=0$.\\
3) for all $|\mu|$ with $\l'<|\mu|<\l$, and all $0\leq j \leq e_{|\mu|}-1$: $\phi_{\pm\mu,j}^-(x_{|\mu|},x_{-|\mu|})=0$.\\
4) for all $d_\l -1 \geq j>k$: $\phi_{\l,j}(x_\l)=0$.\\
5) for all $e_{|\l'|} -1 \geq j>k'$: $\phi_{\pm\l',j}^-(x_{|\l'|},x_{-|\l'|})=0$.\\  
6) $\phi_{\l,k}(x_\l)>0$.\\
7) $\phi_{\pm\l',k'}^-(x_{|\l'|},x_{-|\l'|})>0$.\\
Consider the set $\Delta_U=\{ (\l, \l', k, k') | \l>0 \in Spec(\A),|\l'|<\l\in |Spec(\A)|,  
k\in \{1,\dots,d_\l-1 \},k'\in \{1,\dots,e_{|\l'|}-1 \} \}$.
If $x$ belongs to \begin{equation}\label{eqU}U=\bigvee_{(\l, \l', k, k')\in\Delta_U} U_{k,k'}^{\l,|\l'|},\end{equation} then $x$ is ANT.
\end{prop}

\begin{ex}\label{exU}\emph{(Running example)}:
Again, we illustrate how the conditions are generated for our running Example \ref{exS}. There are only two possible cases to consider.
\begin{small}
\begin{multline*}
U_{0,0}^{2,1}\equiv (x_{2,1}-x_{-2,1}=0)\wedge (x_{1,2}-x_{-1,2}=0)\\\wedge (x_{2,1}>0) \wedge (x_{1,1}+x_{1,2}-x_{-1,1}-x_{-1,2}>0)
\end{multline*}
\vspace*{-5ex}
\begin{multline*}U_{0,1}^{2,1}\equiv (x_{2,1}-x_{-2,1}=0)\wedge (x_{2,1}>0) \wedge (x_{1,2}-x_{-1,2}>0)
\end{multline*}
\vspace*{-3ex}
\end{small}

We obtain $U= U_{0,0}^{2,1} \vee U_{0,1}^{2,1}$.
\end{ex}

Finally, we give a last set of constraints, giving the remaining $ANT$ values. 

\begin{prop}\label{3}
Let $\l$ be a positive eigenvalue and $|\l'|<\l$ in $|Spec(\A)|$ such that:\\
1) $P_{\mu}(x_\mu)$ is zero whenever $|\mu|>\l$,\\
2) $Q_{\pm\l}^+(x_{\pm\l})=0$ and $Q_{\pm\l}^-(x_{\pm\l})=2P_\l(x_\l)$ 
has a positive dominant term,\\
3) $Q_{\pm\mu}^+(x_{\pm\mu})$ is zero whenever $|\l'|<|\mu|<\l$,\\
4( $Q_{\pm\l'}^+(x_{\pm\l'})$ has a positive dominant term.

Then $P(\A,\f)$ is $ANT$ on $x$.
\end{prop}

We now write the preceding proposition in terms of the linear forms $\phi_{\l,j}$, $\phi_{\pm\l,j}^+$ and $\phi_{\pm\l,j}^-$. 

\begin{prop}\label{3'}
For $\l>0$ and $\l'$ in $Spec(\A)$, with $|\l'|<\l$, an integer $k$ between $1$ and $d_\l-1$, and an integer 
$k'$ between $1$ and $d_{\l'}-1$, we denote 
by $V_{k,k'}^{\l,|\l'|}$ the set of $x$ in $E$ which satisfy:\\
1) for all $\mu$ with $|\mu|>\l$, and all $j$ between $0$ and $d_\mu-1$: $\phi_{\mu,j}(x_\mu)=0$.\\
2) for all $0\leq j \leq e_{|\l|}-1 $: $\phi_{\pm\l,j}^+(x_\l,x_{-\l})=0$.\\
3) for all $|\mu|$ with $\l'<|\mu|<\l$, and all $0\leq j \leq e_{|\mu|}-1$: $\phi_{\pm\mu,j}^+(x_{|\mu|},x_{-|\mu|})=0$.\\
4) for all $d_\l -1 \geq j>k$: $\phi_{\l,j}(x_\l)=0$.\\
5) for all $e_{|\l'|} -1 \geq j>k'$: $\phi_{\pm\l',j}^+(x_{|\l'|},x_{-|\l'|})=0$.\\  
6) $\phi_{\l,k}(x_\l)>0$.\\
7) $\phi_{|\l'|,k'}^+(x_{|\l'|},x_{-|\l'|})>0$.\\
Consider the set $\Delta_V=\{ (\l, \l', k, k') |  \l>0 \in Spec(\A),|\l'|<\l\in |Spec(\A)|,  
k\in \{1,\dots,d_\l-1 \},k'\in \{1,\dots,e_{|\l|}-1 \}\}$.
If $x$ belongs to \begin{equation}\label{eqV}V=\bigvee_{(\l, \l', k, k')\in\Delta_V} V_{k,k'}^{\l,|\l'|},\end{equation} then $x$ is ANT.
\end{prop}

\begin{ex}\label{exV} \emph{(Running example)}: 
On our running example, this gives the conditions: 
\begin{small}
\begin{multline*}
V_{0,0}^{2,1}\equiv (x_{2,1}+x_{-2,1}=0)\wedge (x_{1,2}+x_{-1,2}=0)\wedge (x_{2,1}>0)\\\wedge (x_{1,1}+x_{-1,2}+x_{-1,1}+x_{-1,2}>0)
\end{multline*}
\vspace{-5ex}
\begin{multline*}
V_{0,0}^{2,1}\equiv (x_{2,1}+x_{-2,1}=0)\wedge (x_{1,2}+x_{-1,2}=0)\wedge (x_{2,1}>0)\\\wedge (x_{1,1}+x_{-1,2}+x_{-1,1}+x_{-1,2}>0)
\end{multline*}
\end{small}
By  Proposition \ref{3'}, the set $V$ of $ANT$ values is $V= V_{0,0}^{2,1} \cup V_{0,1}^{2,1}$.
\end{ex}

\noindent We now state the main theorem.

\begin{thm}
An element $x$ in $E$ is ANT for $P(\A,\f)$ if and only if we are in the situation of Propositions \ref{1}, \ref{2}, or \ref{3}.
\end{thm}

We can rewrite the preceding theorem.

\begin{thm}\label{SUV}
The set $ANT(P(\A,\f))$, of ANT points of $P(\A,\f)$, is equal to the disjoint union $S\cup U \cup V$, where $S$, $U$ and $V$ are defined in  Propositions \ref{1'}, \ref{2'}, and \ref{3'}. 
\end{thm}

We illustrate the conclusion of Theorem \ref{SUV} next.

\begin{ex}\label{ex-general} \emph{(Running example)}:
Using to Theorem \ref{SUV} the $ANT$ set of our running example is $S \vee U \vee V$ where the sets $S$, $U$ and $V$ have already been explicitly computed.
See Examples \ref{exS}, \ref{exU}, and \ref{exV}.  
Our prototype generates the equivalent semi-linear system:
\begin{scriptsize}
\begin{quote} 
\begin{verbatim}
Locus of ANT
[[[max(-X[-2,1], X[-2,1]) < X[2,1]]]
OR[[X[1,2] == 0, X[-1,2] == 0, X[2,1] == 0, 
X[-2,1] == 0, max(-X[-1,1], X[-1,1]) < X[1,1]]]
OR[[X[1,2] == -X[-1,2], X[2,1] == 0, X[-2,1] == 0, 
-X[-1,1] < X[1,1], X[-1,2] < 0]] 
OR[[X[2,1] == 0, X[-2,1] == 0, 
X[-1,1] - X[1,2] + X[-1,2] < X[1,1], -X[-1,2] < X[1,2]]]
OR[[X[2,1] == 0, X[-2,1] == 0, 
max(-X[-1,2], X[-1,2]) < X[1,2]]]]
OR[[[X[1,2] == X[-1,2], X[2,1] == X[-2,1], 
X[-1,1] < X[1,1], 0 < X[-2,1]]]
OR[[X[2,1] == X[-2,1], X[-1,2] < X[1,2], 0 < X[-2,1]]]]
OR[[[X[1,2] == X[-1,2], X[2,1] == -X[-2,1], 
-X[-1,1] - 2*X[-1,2] < X[1,1], X[-2,1] < 0]]
OR[[X[2,1] == -X[-2,1], -X[-1,2] < X[1,2], X[-2,1] < 0]]]
\end{verbatim}  
\end{quote} 
\end{scriptsize}
\end{ex}

\subsection{The general case}\label{real}

Here we do not suppose that the spaces $K(\A,\f)$ and $E_0(\a)$ are reduced to zero
anymore, but $Spec(\A)$ is still assumed to be real. 
We first make the following definition.

\begin{df}
For $x$ in $E$, we denote by 
$E(\A,x)$ the subspace of $E$ generated by the family $(\A^k(x))_{k\geq 0}$. It is an $\A$-stable subspace.
\end{df}

We next give the asymptotic behavior of 
$\f(\A^k (x))$ for $k$ large, $x\in E_{\l}(\A)$, and $\l \in Spec(\A)-\{0\}$, of which Proposition \ref{version1} was a special precise case.

\begin{prop}\label{polynomial}
For $\l \in Spec(\A)-\{0\}$, and $x$ in $E_\l(\A)$, there exists $P_\l(\f,x) \in \R[T]$ such that $\f(\A^k( x))=\l^k P(\f,x)(k)$. 
\end{prop}

In this situation, we will write $P_\l(v,x,k)$ for $P_\l(v,x)(k)$, so that $P_\l$ is a map from $\R^n\times E_\l(\A)\times \N$ to $\R$, 
linear in the first two variables, and polynomial in the last. We have the following:

\begin{prop}
If $\l\neq 0$ is a real eigenvalue of $\A$, $\f$ belongs to $E^*$, and $x$ belongs to $E_\l(\A)$, then $P_\l(\f,x,.)$ 
is nonzero if and only if $x\notin K(\f,\A)$.
\end{prop}

When studying the locus of $ANT$ values in $E$, or  for any question related 
to the termination of  program $P(\A,\f)$, the subspace $K(\f,\A)$ is not important since we have the following:.

\begin{prop}
For any $k\geq 0$, the linear form $\f\circ \A^k$ factors through the quotient $E/K(\A,\f)$, i.e., for any $x \in E$, the value 
of $\f(\A^k(x))$  depends only on the class $x+K(\A,\f)$.
\end{prop}

We denote by $\overline{\A}$ the endomorphism of 
$\overline{E}=E/K(\A,\f)$ induced by $\A$, and by $\overline{\f}$ the linear form on $E/K(\A,\f)$ induced by $\f$. Then, we 
write $\overline{E}=\overline{E}_0(\overline{\A})\oplus \overline{E}^a$, where 
$\overline{E}^a=\oplus_{\l\in Spec(\overline{\A})-\{0\}}\overline{E}_\l(\overline{\A})$.
Consider the restriction $\overline{\A}^a$ 
of $\overline{\A}$ to $\overline{E}^a$, as well as the restriction $\overline{\f}^a$ 
of $\overline{\f}$ to $\overline{E}^a$. Program $P(\overline{\A}^a,\overline{\f}^a)$ is of the form studied in the previous section, 
that is, we have $\overline{E}^a_0(\overline{\A}^a)=K(\overline{\A}^a,\overline{\f}^a)=\{0\}$. Hence, we know how to compute 
$ANT(P(\overline{\A}^a,\overline{\f}^a))$.  The main theorem of this section  reduces the $ANT$ computation of $P(\A,\f)$ to that of $ANT(P(\overline{\A}^a,\overline{\f}^a))$.

\begin{thm}\label{reductiontoregular}
Program $P(\A,\f)$ terminates if and only if  program $P(\overline{\A}^a,\overline{\f}^a)$ terminates. 
Moreover, if we write the canonical projection $p:E\rightarrow \overline{E}$, we have the relation 
$ANT(P(\A,\f))=p^{-1}(ANT(P(\overline{\A}^a,\overline{\f}^a))+\overline{E}_0(\overline{\A}))$.
\end{thm}

It might not be obvious  how to apply this in a concrete situation, 
where we are given a program $P(A,f)$, corresponding to a matrix $A\in M(n,\R)$, and a row vector $f$  in $\R^n$. 
We explain how to proceed.  First, compute a basis $B_{A,f}$ of  $K(A,f)=\cap_{k=0}^{n-1} Ker(fA^k)$, which is the kernel of the matrix  of $M(n,\R)$ with its $i$-th row equal to $fA^{i-1}$.  Then, take any family $B_1$ where $B'=B_{A,v}\cup B_1$ is a basis of $\R^n$, and let $P$ be the  matrix whose columns are the vectors of $B'$. 
We have $P^{-1}AP= \begin{small}\begin{pmatrix} X & Y \\ 0 & A_1 \end{pmatrix}\end{small}$, for $A_1$ the size  of $B_1$. Now consider the matrix $A_1\in M(n_1,\R)$, and take the modified Jordan basis $B_J$ where the first vectors of $B_J$ are a Jordan basis $B_0$ of $E_0(A_1)$, and the next vectors in $B_J$ are ordered as a union of basis $B_\l$ for each $E_\l(A_1)$, with $\l\neq 0$ in $Spec(A_1)$, where $B_\l$ is the modified Jordan basis defined in Lemma \ref{goodbasis}.  
If $P_1$ is the matrix in $M(n_1,\R)$, whose columns are the vectors of $B_J$, then $T=P_1^{-1}A_1P_1$ is of the form 
$T=diag(T_0,\l_1 T_{\l_1},\dots,\l_{t-1} T_{\l_{t-1}},\l_t T_{\l_t})$ where 
${\l_1,\dots,\l_t}$ are the nonzero eigenvalues of $A_1$, $T_{\l_i}$ is of the form described in Lemma \ref{goodbasis}, and 
$T_0$ is of the form $\begin{scriptsize}\begin{pmatrix}  0 & 1 & & & \\ & 0 & 1 & & \\  & & \ddots & \ddots & \\
 &  &  & 0 & 1 \\   &  &  &  & 0 
\end{pmatrix}\end{scriptsize}$.  Write $T^a=diag(\l_1 T_{\l_1},\dots,\l_t T_{\l_t})$, so that 
$T=diag(T_0,T^a)$ in $M(n_a,\R)$, where $n_a=\sum_{\l \neq 0 \in Spec(A_1)} dim(E_\l(A_1))$.
If we write $Q=diag(I_{n-n_1},P_1)$, and $R=PQ$, we get
\begin{small}$B=R^{-1}AR= \begin{pmatrix} X & Y \\ 0 & T \end{pmatrix}= \begin{pmatrix} X & Y \\ 0 & diag( T_0 , T^a) \end{pmatrix}$\end{small}.
For $x=(x_1,\dots,x_n)^T\in \R^n$, we write $x^a= (x_{n-n_a+1},\dots ,x_n)^T$. 
Then, we set $w=fR$ in $M(1,n,\R)$, and write $w^a=(w_{n-n_a+1},\dots,w_{n})$. We know how to compute the set $ANT(P(T^a,w^a))$using the results of the previous section.
We finally obtain the following theorem.

\begin{thm}\label{thmE0K}
Vector $x$ is in  $ANT(P(B,w))$ if and only if $x^a$ is in $ANT(P(T^a,w^a))$. Vector 
$y$ is in $ANT(A,f)$ if and only if $R^{-1}y$ is in $ANT(P(B,w))$, i.e., $ANT(P(A,v))=R (ANT(P(B,w)))$. In particular, 
 $P(A,f)$ terminates if and only if  $P(T^a,w^a)$ does.
\end{thm}

\begin{ex}
Take a program $P(\A,\f)$ and matrices $Mat_C(\A)=A$ and $Mat_C(\f)=v$ corresponding, respectively, to the linear forms $\A$ and $\f$, expressed in the canonical basis $C$ of $\R^n$:

\begin{scriptsize}$A=\begin{pmatrix}  
1 &   0 &  0 &  0 &  0 &  0 &  0 &  0\\
2 &  0 &  0 &  0 &  0 &  0 &  0 &  0\\
6 & -2 & -1 &  1 &  0 &  0 &  0 &  0\\
10  &-3  &-4  & 3  & 0  & 0  & 0  & 0\\
30 & -15 &  -6 &   7&    0&   -1&   0&   0\\
44 &-28 & -6  & 9  & 1 & -2  & 0  & 0\\
90 & -55 & -9 & 12&   4&  -7&   2&   0\\
57 &-19&  -9&   1&   5&  -8&   4&  -2
\end{pmatrix}$\\ 
and $v=(-1, -2,  1,  0,  0,  0,  0,  1)$\end{scriptsize}.

The main step is construction of a basis $B_{E_0,K}$ in which the matrices of $\A$ and $\f$ are the form $B$ and $w$. 
Follow the steps described above, we obtain the following matrices $R$,   $Mat_{B_{E_0,K}}(\A)=B$, and $Mat_{B_{E_0,K}}(\f)=w$:

\begin{scriptsize}
$R = \begin{pmatrix}
1& 0& 0& 0& 0& 0& 0& 0\\
2& 1& 0& 0& 0& 0& 0& 0\\
3& 3& 1& 0& 0& 0& 0& 0\\
4& 5& 2& 1& 0& 0& 0& 0\\
6& 5& 3& 2& 1& 0& 0& 0\\
4& 2& 5& 2& 1& 1& 0& 0\\
3& 8& 8& 2& 1& 2& 1& 0\\
2& 0& 0& 1& 1& 1& 1& 1 \end{pmatrix}$,\\
  $B=\begin{pmatrix} 1 & 0 & 0 & 0 & 0 & 0 & 0 & 0 \\
0 & 0 & 0 & 0 & 0 & 0 & 0 & 0 \\
0 & 0 & 1 & 1 & 0 & 0 & 0 & 0 \\
0 & 0 & 0 & 1 & 0 & 0 & 0 & 0 \\
0 & 0 & 0 & 0 & -1 & -1 & 0 & 0 \\
0 & 0 & 0 & 0 & 0 & -1 & 0 & 0 \\
0 & 0 & 0 & 0 & 0 & 0 & 2 & 0 \\
0 & 0 & 0 & 0 & 0 & 0 & 0 & -2                
\end{pmatrix}$, and $w=\begin{pmatrix} 0\\ 1\\ 1\\ 1\\ 1\\ 1\\ 1\\ 1\end{pmatrix}$\end{scriptsize}.

Matrix $R$ is such that $B=R^{-1}AR=\begin{small}\begin{pmatrix} X & Y \\ 0 &  \begin{pmatrix} T_0 & \\ & T^a\end{pmatrix} \end{pmatrix}\end{small}$.  
In this case,  $B$ has the expected form with $X=(1)$, $Y=(0\ 0\  0\  0\  0\  0)$, $T_0= (0)$, and $T^a$ being the matrix $T$ depicted in Example \ref{exS}.
Also, if we denote by $e_1$ and $e_2$ the two first elements of the canonical basis $C$, we obtain $K(A,f)=R(K(B,w))=Vect(R(e_1))$, \emph{i.e.},  $Vect(R(e_1))$ is the space spanned by the first column of $R$, and $E_0(B)=Vect(e_2)$. 
By construction
$w=vR=(0,1,1,1,1,1,1,1)$, and thus $w^a=(1,1,1,1,1,1)$.  Finally, we apply Theorem \ref{thmE0K}, which claims the following equivalence:
($y$ is ANT for $P(A,v)$) $\Leftrightarrow$ ($x=(x_1, \dots x_8)^{\top}= R^{-1}y$ is ANT for $P(B,w)$) $\Leftrightarrow$ ($x^a=(x_3,\dots, x_8)^{\top}$ is ANT for $P(T^a,w^a)$). The analysis of the $ANT$ set for $P(A,f)$ is reduced to the generation of $ANT(P(T^a,w^a))$. As $T^a$ and $w^a$ describe the same system as the one obtained in Example \ref{ex-general}, we already have the symbolic representation of $ANT(P(T^a,w^a))$.
To  generate the $ANT$ set for $P(A,v)$, one just needs to rewrite the semi-linear space obtained in Example \ref{ex-general}, now considering the variables $(x_3, \dots , x_8)$.
\end{ex} 

\section{ANT sets generation in practice}\label{practice}
We provide more practical details on our computational method.  We show that in practice one can obtain the $ANT$ set using only a few specific and concise formulas \emph{e.g.},  Theorem \ref{formula} and Equations \ref{1}, \ref{2}, and \ref{3}. 
First we identify a useful characteristic of almost all affine programs. 
Previously, we identified and treated separately and completely the degenerate
cases where the spaces $K(\A,\f)=\cap_{k\geq 0} Ker(\f\circ
\A^k)=\cap_{k= 0}^{n-1} Ker(\f\circ \A^k)$ and $E_0(\A)$ are not
reduced to $\{0\}$.  That is why we made the assumption that $K(\A,\f)=
\{0\}$ and $E_0(\A)=\{0\}$ for the remainder of the paper.

\begin{df}\label{normal}
Let $(A,f)$ belong to $\mathcal{M}(n,\R)\times \R^n$. We say that a program $P(A,f)$ is \emph{normal} if $Spec(A)$ does not contain a real eigenvalue 
and its additive inverse.  In other words, if the $\l\in Spec(A)$ then $-\l$ is not an eigenvalue of $A$.
\end{df}

We are going to show that almost all programs are normal. We recall the definition of a Zariski open subset of a real vector space.

\begin{df}
let $V$ be a finite-dimensional vector space over $\R$, and let $P$ in $\R[V]$,
that is, a polynomial map from $V$ to $K$. If $P_1,\dots, P_t$ are in 
$K[V]$, we denote by $D_{P_1,\dots,P_t}=\{v\in V, \exists i\in [1,\dots,t], P_i(v)\neq 0\}$. A Zariski open 
subset of $V$ is a finite intersection of sets of the form $D_{P_1,\dots,P_t}$.\qed
\end{df}

The following lemma is standard.

\begin{LM}
A non-empty Zariski open subset of $V$ is open, dense, and its complementary set in $V$ has  zero Lebesgue measure. Two non-empty 
Zariski open sets have a non-empty Zariski open intersection.
\end{LM}

We can now state and prove the main theorem of this section.

\begin{thm}\label{Zar}
The set $R(A,v)$ of pairs $(A,v)$ with $A$ in $\mathcal{M}(n,\R)$ and $v$ in $\R^n$, where $A$ has no real eigenvalue $\l$ with $-\l$ is also an eigenvalue, and such that the space $K(A,v)$ is reduced to zero, contains a non empty Zariski open subset of $\mathcal{M}(n,_R)\times \R^n$. In particular, its complementary set in $\mathcal{M}(n,\R)\times \R^n$ has  zero Lebesgue measure.
This basically says, that a program $P(A,v)$ is almost always regular.\qed
\end{thm}

We now present the practical details of our procedure for generating the formulas that compose the symbolic representations of the $ANT$ set for normal programs.  We first recall the fact that one can  produce a basis $B_\l$ of $E_\l(\A)$ such that $Mat_{B_\l}(\A)$ is of the form $\l. diag(T_{\l,1},\dots,T_{\l,r_\l})$, where each $T_{\l,i}$ is a matrix of the form depicted in Lemma \ref{goodbasis}.  The power of $T_\l$ is indicated in Lemma \ref{poweroflambda}.  Again, we write $B=B_{\l_1} \cup \dots \cup B_{\l_t}$ a basis of $E$, and $Mat_B(f)$ as  $(a_{\l_1,1},a_{\l_1,2},\dots,a_{\l_1,d_{\l_1}},\dots,a_{\l_t,1},\dots,a_{\l_t,d_{\l_t}})$. For $\l$ in $Spec(\A)$ let $P_{\l,j}=a_{\l,1} (_{j-1}^{k})+a_{\l,2} (_{j-2}^{k})+\dots+a_{\l,j}$, as in Proposition \ref{version1}, for $j$ between $1$ and $d_{\l}$, and such that $Mat_B(\f\circ \A^k)=(P_{\l,1}(k),\dots,P_{\l,d_{\l}}(k))$. When $x=\sum_{\l\in Spec(\A)} x_\l$, we write $P_\l(x_\l,k)=\sum_{j=1}^{d_\l} x_{\l,j} P_{\l,j}(k)$.

\begin{thm}\label{thm-algo}
Suppose that we are in the common situation 
where $\A$ has no eigenvalue $\l$ with $-\l$ is also an eigenvalue.
Then $x$ is $ANT^r$ if and only if the following two conditions hold:
\begin{enumerate}
\item The $\l$ of highest absolute value satisfying : $\exists j \in \{1, \dots d_{\l}\}$ such that $a_{\l , j}x_{\l,j}\neq 0$ is strictly positive. 
\item For this $\l$, the highest $j_0\in \{1, \dots d_{\l}\}$ such that $a_{\l , j_0}x_{\l , j_0}\neq 0$ satisfies $a_{\l , j_0}x_{\l , j_0}>0$. \qed
\end{enumerate}
\end{thm}

Now we use theorem \ref{thm-algo} to get the generic formula that, once instantiated,  represent the $ANT$ set symbolically and exactly.

\begin{thm}\label{formula}
For $\l > 0$ in $Spec(A)$ and an integer $k$ between $1$ and $d_{\l}$, denote by $S_{\l, k}$ the set of $x$ in $E$ which satisfy:\\
\vspace*{-2ex} 
\begin{multline}\label{1}
\text{If $\mu \in Spec(\A)$ is such that}\text{ $|\mu|>\l$ then}\\ \,\,\forall h \in\{1, \dots ,d_{\mu}\}:
a_{\mu , h}x_{\mu , h}=0.\end{multline}
\begin{equation}\label{2}\forall h \in\{k+1, \dots ,d_{\l}\}:\ \ a_{\l , h}x_{\l , h}=0.\end{equation}
\begin{equation}\label{3} a_{\l , k}x_{\l , k}>0.\end{equation}  
Consider the set $\Delta_S=\{(\l, k)| \l>0 \in Spec(\A), k\in \{1,\dots,d_\l \}\}$.  Then we have \begin{equation}ANT(P(A, f))=\bigvee_{(\l, k)\in\Delta_S} S_{\l,k}.\qed\end{equation}\label{5}
\end{thm}

When using Theorem \ref{formula}, one needs first to evaluate the terms $ a_{\l , k}$ and $x_{\l , k}$. As we obtain our results using the decomposition of $x$ and $f$ in $B$, we recall in the following lemma how one obtains it from the decomposition of $x$ and $f$ in $B_c$, the canonical basis.
 
\begin{LM}\label{p-1u}
Let $P$ be the transformation matrix corresponding to $B$, and
$x\in E$. If $x=\sum_{i=1}^n x_i e_i= (x_1,..., x_n)^{\top}\in B_c$,
and $x$ decomposes as $\sum_{j=1}^t (\sum_{i=1}^{d_j} x_{\l_j,i}
e_{\l_j,i})$ in $B$, then the coefficients $x_{\l_j,i}$ are those of the
column vector $P^{-1} x$ in $B_c$.\qed
\end{LM}

This lemma is illustrated in the first computational step presented next.
We first provide an example showing the main steps when computing $ANT$ sets for normal programs.
\begin{ex}\label{normal}
Consider the program $P(A,f)$ depicted as follows:
\begin{small}
$A=\begin{pmatrix} 
26&   2& -15&  -6&  30\\
24&   3& -12&  -6&  48\\
32&   0&  -9&   2&  66\\
-12&   1&   6&   8& -24\\
-4&  -1&   3&   0&   0 \end{pmatrix}$, and $f=\begin{pmatrix} -2\\ 0\\-1\\0\\-1/2\end{pmatrix}$\end{small}.  

\noindent\emph{\textbf{Step $1$}}: In the triangularization of matrix $A$ we get:\\
\begin{small}
$P =\begin{pmatrix} 
1&        1&    1&    1&    0\\
4/5&   -3/2&  6/5&    0&    1\\
1&        0&  8/5&  2/3&  2/3\\
-2/5&   1/2& -3/5&    0& -1/2\\
-1/5&  -1/2& -1/5& -1/3&  1/6
 \end{pmatrix}$,\\ 
$D =
\begin{pmatrix} 
9& 0& 0& 0& 0\\
0& 5& 0& 0& 0\\
0& 0& 2& 0& 0\\
0& 0& 0& 6& 0\\
0& 0& 0& 0& 6
\end{pmatrix},$ and\\
$P^{-1} =
\begin{pmatrix}  0&   0&  -5& -10& -10\\
0&  -2&   0&  -4&   0\\
-5&   0&   0&  -5& -15\\
6&   2&   5&  19&  25\\
6&  -2&   4&   8&  26 \end{pmatrix}$.\end{small}

We obtain the following eigenvectors given by the column of  $P$, written using our notation:\\
\begin{small}
$e_{9, 1} = (1, 4/5, 1, -2/5, -1/5)^{\top}$,\\
$e_{5, 1} = (1, -3/2,0,1/2,-1/2)^{\top}$,\\ 
$e_{2, 1} = (1, 6/5,8/5,-3/5,-1/5)^{\top}$,\\ 
$e_{6, 1} = (1, 0, 2/3, 0,-1/3)^{\top}$ and\\ 
$e_{6, 2} = (0, 1,2/3,-1/2,1/6)^{\top}$.\\
\end{small}
\noindent\emph{\textbf{Step $2$}}: Computing $S_{\l, k}$ for all positive $\l \in Spec(A)^*$ and $k\in\{1,\dots , d_{\l}\}$:
\begin{itemize}
\item Our algorithm first computes the coefficients $a_{\l,i}$:\\
\begin{scriptsize}
\vspace*{-4ex}
\begin{multline}
a_{9,1}= <f,e_{9,1}>\\=<(-2,0,-1,0,-1/2)^{\top},(1, 4/5, 1, -2/5, -1/5)^\top\\>=-29/10,
\end{multline}
\vspace*{-5ex}
\begin{multline}
a_{5,1}= <f,e_{5,1}>\\=<(-2,0,-1,0,-1/2)^{\top},(1, -3/2,0,1/2,-1/2)^\top\\>=-7/4,
\end{multline}
\vspace*{-5ex}
\begin{multline}
a_{2,1}= <f,e_{2,1}>\\=<(-2,0,-1,0,-1/2)^{\top},(1, 6/5,8/5,-3/5,-1/5)^\top\\>=-7/2,
\end{multline}
\vspace*{-5ex}
\begin{multline}
a_{6,1}= <f,e_{6,1}>\\=<(-2,0,-1,0,-1/2)^{\top},(1, 0, 2/3, 0,-1/3)^\top\\>=-5/2,
\end{multline}
\vspace*{-5ex}
\begin{multline}
a_{6,2}= <f,e_{6,2}>\\=<(-2,0,-1,0,-1/2)^{\top},(0, 1,2/3,-1/2,1/6)^\top>\\=-3/4.
\end{multline}
\end{scriptsize}

\item Now our algorithm computes the coefficients of the decomposition of the initial variable values in $B$. 
They are  the column vector $P^{-1}\cdot u$ in $B_c$ where $u=(u_1, u_2, u_3, u_4, u_5)^\top$ is the vector encoding the initial variable values.\\ 
\begin{scriptsize}
$P^{-1}.\begin{pmatrix} u_1  \\
u_2  \\
u_3  \\
u_4  \\
u_5  
 \end{pmatrix} = \begin{pmatrix} 
-5*u3 - 10*u4 - 10*u5\\
-2*u2 - 4*u4\\
-5*u1 - 5*u4 - 15*u5\\
6*u1 + 2*u2 + 5*u3 + 19*u4 + 25*u5\\
6*u1 - 2*u2 + 4*u3 + 8*u4 + 26*u5]
 \end{pmatrix} = \begin{pmatrix} x_{9,1}  \\
x_{5,1}  \\
x_{2,1}  \\
x_{6,1}  \\
x_{6,2}
 \end{pmatrix}$.\end{scriptsize}\\
\end{itemize}
\noindent\emph{\textbf{Step $3$}}: We apply Theorem~\ref{formula} to generate the ANT Locus.
In order to computes the sets $S_{\l, i}$ one needs to first generate the formulas appearing in 
Theorem~\ref{formula} for each positive eigenvalue. 
Then one needs to instantiate the generic formulas according to the computed terms $a_{\l, i}$s and $x_{\l, j}$s.  
In the following, we show all computational operations in this example.  
For each positive eigenvalue $\l$ and integer $k\in\{1, ..., d_{\l}\}$, we directly apply Theorem \ref{formula}, using its three generic formulas.

\begin{itemize}

\item Case \textbf{$\l=9$} and \textbf{$k=1$}:
Eqs. \ref{1} and \ref{2}, from  Theorem \ref{formula}, induce no constraint in this case.
Firstly because there is no $\mu \in Spec(\A)$ such that $|\mu|>\l$, and secondly because $\{k+1, ..., d_{\l} \}=\emptyset$ since $k=d_{\l}=1$. Eq, \ref{3}, from Theorem \ref{formula}, generates the constraint $S_{9, 1}=(a_{9,1}x_{9,1}>0)$. \\  

\item Case \textbf{$\l=5$} and \textbf{$k=1$}:
Eq. \ref{2} from  Theorem \ref{formula} induce no formula since $\{k+1, ..., d_{\l} \}=\emptyset$, as we are in the case where $k=d_{\l}=1$.  
Considering Eq. \ref{1}, one needs to treat the two sub-cases with $\mu=9$ and $\mu=6$. When $\mu=9$, we get  $a_{9,1}x_{9,1}=0$, and when $\mu=6$, we obtain  $(a_{6,1}x_{6,1}=0)\wedge (a_{6,2}x_{6,2}=0)$. 
Eq. 3 induces the formula $a_{5,1}x_{5,1}>0$, and we generate the following equation associated to this eigenvalue: 
\begin{multline}
S_{5,1}= (a_{9,1}x_{9,1}=0)\wedge (a_{6,1}x_{6,1}=0)\wedge (a_{6,2}x_{6,2}=0)\\\wedge (a_{5,1}x_{5,1}>0).
\end{multline}

\item Case \textbf{$\l=2$} and \textbf{$k=1$}:
With Eq. \ref{1}, we have three sub-cases when $\mu=9$, $\mu=6$, and when $\mu=5$.  Starting with  $\mu=9$, we generate the constraint $a_{9,1}x_{9,1}=0$.  
With $\mu=6$, we obtain $(a_{6,1}x_{6,1}=0)\wedge (a_{6,2}x_{6,2}=0)$, and when $\mu=5$ we have  $a_{5,1}x_{5,1}=0$. 
Here, Eq. \ref{2} generates no further formulas since $\{k+1, ..., d_{\l} \}=\emptyset$, as we have $k=d_{\l}=1$. With Eq. \ref{3}, we obtain the constraint $(a_{2,1}x_{2,1}>0)$ and the formula 
\begin{multline}
S_{2,1}= (a_{9,1}x_{9,1}=0)\wedge (a_{6,1}x_{6,1}=0)\wedge (a_{6,2}x_{6,2}=0)\\\wedge (a_{5,1}x_{5,1}=0) \wedge (a_{2,1}x_{2,1}>0).
\end{multline}

\item Case \textbf{$\l=6$} and \textbf{$k=1$}:
With Eq. \ref{1}, one needs to consider $\mu=9$, which gives the formula $(a_{9,1}x_{9,1}=0)$. 
Now, looking at Eq. \ref{2}, we have $h=2$ and we obtain the constraint $(a_{2,2}x_{2,2}=0)$.   
With Eq. \ref{3} we obtain the constraint $(a_{2,1}x_{2,1}>0)$, and  generate the formula
$$S_{6,1}= (a_{9,1}x_{9,1}=0)\wedge (a_{2,2}x_{2,2}=0)\wedge(a_{2,1}x_{2,1}>0).$$

\item Case \textbf{$\l=6$} and \textbf{$k=2$}:
Again, with Eq. \ref{1} one needs to consider $\mu=9$ which gives the formula $(a_{9,1}x_{9,1}=0)$.   
Also, Eq. \ref{2} induces no formula since $\{k+1, ..., d_{\l} \}=\emptyset$ because 
we have  $k=d_{\l}=2$.  
Using Eq. \ref{3}, we get the constraint $(a_{2,2}x_{2,2}>0)$, and generate the formula:
$$S_{6,2}= (a_{9,1}x_{9,1}=0)\wedge (a_{2,2}x_{2,2}>0).$$
 
\end{itemize}

According to Theorem \ref{formula}, Eq. \ref{5}, the $ANT$ locus $S$ reduces to the following semi-linear space:
$$S= S_{9,1}\vee S_{5,1} \vee S_{2,1} \vee S_{6,1} \vee S_{6,2}.$$
The initial values of $x, y, z, s, t$ are represented, respectively, by the parameters $u_1, u_2, u_3, u_4, u_5$. Now, we can express the results in the canonical basis 
 using  Lemma \ref{p-1u} if we want, as all the terms $a_{\l, i}$ and $x_{\l,j}$ have been already computed in Step 1.\qed

\end{ex}

The pseudo code depicted in Algorithm \ref{Algo-1} illustrates the
strategy.  Our algorithm takes as input the number of variables, the
chosen field where the variables are interpreted, the assignment
matrix $A$ and the vector $f$ encoding the loop condition.  We first
compute the list of positive eigenvalues. See lines $1$ and $2$ in
\ref{Algo-1}.  Then, we just need to directly encode the statements
and formulas provided in Theorem \ref{formula}.  We proceed
considering each positive eigenvalues $e'[i]$ at a time, ee line $3$, for
each $k$ in $\{1, ..., d_\l\}$, see line $5$.
\begin{itemize}
\item Then, we generate the constraint given by equation \ref{1}. We
  look for $\mu \in Spec(\A)$ such that $|\mu|>\l$, see line $7$. Then,
  for all $h \in\{1, \dots ,d_{\mu}\}$, see  line $10$, we add the
  constraint $(a_{\mu , h}x_{\mu , h}=0)$ indicated in line $11$.
\item Next, we consider Eq. \ref{2}, in line $12$, and we add the constraint $\ a_{\l , h}x_{\l , h}=0$, as in line $13$.
\item Proceeding, we consider Eq. \ref{3} and we add the constraint $( a_{\l , k}x_{\l , k}>0)$, by line $14$.
\end{itemize}
Finally, we progressively compute the disjunction $\bigvee_{(\l, k)\in\Delta_S} S_{\l,k}$, as in line $15$.

\begin{small}
\begin{algorithm}
{\bf /*Generating the $ANT$ set.*/}\;
\KwData{$n$ the number of program variables, $\mathbb{K}$ the field, $P(A,f)\in P^{\H}$ where $A\in\mathcal{M}(n,\mathbb{K})$ and $f\in\mathcal{M}(n,1,\mathbb{K})$}
\KwResult{$ANT^r(P(A,f))$}
\Begin{
\nl $\{e[1],...,e[r]\} \longleftarrow$ {\bf eigenvalues(}$A${\bf)}\;
\nl $\{e'[1],...,e'[s]\} \longleftarrow$ {\bf strictly\_positives(}$\{e[1],...,e[r]\}${\bf)}\; 
\nl \For{$i=1$ \KwTo $s$}{       
         \nl $d_{\l} \longleftarrow$ {\bf multiplicity(}$e'[i]${\bf )}\;
   \nl \For{$k=1$ \KwTo $d_{\l}$}{
         \nl \For{$p=1$ \KwTo $r$}{
              \nl \If {$|e[p]|>e'[i]$} {
\nl $\mu \longleftarrow e[p]$\;
\nl $d_{\mu} \longleftarrow$ {\bf multiplicity(}$e[p]${\bf )}\;
                   \nl \For{$h=1$ \KwTo $d_{\mu}$}{
                    \nl $Ant[i]\longleftarrow Ant[i] \wedge (a_{\mu, h}x_{\mu, h}=0)$\;
                                               } 
                                       }
                                  }    
                  \nl \For{$l=k+1$ \KwTo $d_{\l}$}{
                  \nl $Ant[i]\longleftarrow Ant[i] \wedge (a_{e'[i], l}x_{e'[i], l}=0)$\;
                                                }
                  \nl $Ant[i]\longleftarrow Ant[i] \wedge (a_{e'[i], k}x_{e'[i], k}>0)$\;
         \nl $ANT\longleftarrow ANT \vee Ant[i]$\;         
         }
}
     \nl    \Return $ANT$\;                                     
}
\caption{  {\bf ANT\_linear\_Loop} $(n,\mathbb{K}, A, f${\bf )}\label{Algo-1}}
\end{algorithm}
\end{small}

\section{ANT Algorithm and Experiments}\label{experiments}

In Table \ref{tab-experim} we list some experimental results. The column \textbf{Set-i} refers to a set of loops generated randomly.
As expected,  the probability to produce terminating programs tends to zero when the number of variables grows. The column \textbf{\#Loops} gives the number of loops treated, where each set includes the analysis of $500$ loops.  The column \textbf{Class} gives the class of the linear loop programs either. 
$P^{\H}$, $P^{\G}$ or $P^{\mathbb{A}}$.  The column \textbf{\#Cond} gives the number of conjunctions in the loop condition for each program, and \textbf{\#Var} refers to the numbers of program variables. The column \textbf{\#T} returns the number of terminating programs, and the column  \textbf{\#NT} gives the number of non-terminating programs. Finally,  column \textbf{CPU/s[ANT]} gives the cpu time for deciding on termination and the computation of the $ANT$ loci.
We have implemented our prototype using \textbf{Sage} \cite{sage} using  interfaces written in python.  
This way we had access to several  mathematical packages that were used to guarantee that all random loops were triangulable in the corresponding field, even when we had lots of variables.   

\begin{ex} Here we show an example of $ANT$ computations taken from the long output results refering to line $7$ (with $10$ variables):

\noindent\begin{scriptsize}
\begin{verbatim}

=====Vector F====
[  1   0   1   0 1/2 1/2   0   2  -1   2]
=====Matrix A====
[ -67   55   -1   19   15   -5   12   -4 -340  -81]
[ -15 -132  -27  -15  -15  -15  -12  -12  357  462]
[ -36  -34  -13    2    0  -10    0   -8   10  170]
[-124   28  -20   37   20  -20   16  -16 -364   72]
[ 111   -4   22  -22   -7   20   -8   16  271 -127]
[  -2 -147  -29  -21  -20  -12  -16  -12  423  485]
[  36   34   14   -2    0   10    4    8  -10 -164]
[  20  164   36   20   20   20   16   20 -424 -564]
[  13  -24   -2   -6   -5    0   -4    0  105   59]
[ -18  -17   -7    1    0   -5    0   -4    5   86]
=================
Locus of ANT:[-20*u10 + 5*u4 + 20*u9 > 0, 6*u1 - 6*u10 + 18*u9 == 0]
OR[22*u10 - 11*u3 > 0, -20*u10 + 5*u4 + 20*u9 == 0, 123/2*u10 - 41/2*u2 + 123/2*u9 == 0,
6*u1 - 6*u10 + 18*u9 == 0, 7/2*u1 + 19*u10 + 7/2*u2 + 7/2*u3 + 7/2*u4 + 9*u5 + 7/2*u6 + 9*u7 + 7/2*u8 - 13*u9 == 0,
-24*u10 - 9*u7 - 3/2*u8 - 6*u9 == 0, -17/2*u1 - 101/2*u10 + 17*u2 + 17/2*u3 - 17/2*u4 - 17/2*u5 - 3*u6 - 163/2*u9 == 0]
OR[7/2*u1 + 19*u10 + 7/2*u2 + 7/2*u3 + 7/2*u4 + 9*u5 + 7/2*u6 + 9*u7 + 7/2*u8 - 13*u9 > 0,
-20*u10 + 5*u4 + 20*u9 == 0, 123/2*u10 - 41/2*u2 + 123/2*u9 == 0, 6*u1 - 6*u10 + 18*u9 == 0]OR 
[-24*u10 - 9*u7 - 3/2*u8 - 6*u9 > 0, -20*u10 + 5*u4 + 20*u9 == 0, 123/2*u10 - 41/2*u2 + 123/2*u9 == 0,
6*u1 - 6*u10 + 18*u9 == 0, 7/2*u1 + 19*u10 + 7/2*u2 + 7/2*u3 + 7/2*u4 + 9*u5 + 7/2*u6 + 9*u7 + 7/2*u8 - 13*u9 == 0]
OR[-17/2*u1 - 101/2*u10 + 17*u2 + 17/2*u3 - 17/2*u4 - 17/2*u5 - 3*u6 - 163/2*u9 > 0,
-20*u10 + 5*u4 + 20*u9 == 0, 123/2*u10 - 41/2*u2 + 123/2*u9 == 0, 6*u1 - 6*u10 + 18*u9 == 0,
7/2*u1 + 19*u10 + 7/2*u2 + 7/2*u3 + 7/2*u4 + 9*u5 + 7/2*u6 + 9*u7 + 7/2*u8 - 13*u9 == 0,
-24*u10 - 9*u7 - 3/2*u8 - 6*u9 == 0] [0.05]
\end{verbatim}  
\end{scriptsize}

\end{ex}

\begin{table}[t]
\caption{Experiments on randomly generated linear loop programs}\label{tab1} 
\begin{center}
\begin{scriptsize}
\begin{tabular}{|c|c|c|c|c|c|c|c|c|}
\hline 
       \#Loops& Class & \#Cond & \#Var & \#T & \#NT & CPU/s[ANT] \\ 
\hline $1000$ & $P^{\H}$ &  $1$ & $[3,4]$ & $130$ & $870$ & $19,91$\\ 
\hline $1000$ & $P^{\G}$ &  $[2,4]$ & $[3,4]$ & $125$ & $875$ & $23,72$\\ 
\hline $1000$ & $P^{\A}$ &  $[2,4]$ & $[3,4]$ & $117$ & $883$ & $24,57$\\ 
\hline $1000$ & $P^{\H}$ &  $1$ & $[5,6]$ & $58$ & $942$ &$39.03$\\ 
\hline $1000$ & $P^{\G}$ &  $[2,4]$ & $[4,6]$ & $55$ & $945$ &$45,45$\\ 
\hline $1000$ & $P^{\A}$ &  $[2,4]$ & $[4,6]$ & $52$ & $948$ &$49.79$\\ 
\hline $1000$ & $P^{\H}$ & $1$ & $[7,15]$ & $26$ & $974$ & $107.92$ \\ 
\hline $1000$ & $P^{\G}$ & $2$ & $[7,15]$ & $44$ &  $956$ & $178,08$\\ 
\hline $1000$ & $P^{\A}$ & $2$ & $[7,15]$ & $21$ & $979$  & $187,78$\\ 
\hline 
\end{tabular}
\end{scriptsize}
\label{tab-experim}
\end{center}
\end{table}

\begin{wrapfigure}{o}{3cm}
 \vspace{-10pt}
\begin{scriptsize}
\begin{lstlisting}
y=0;
if (x>=0){
 while(-x >
    -2^(30)){
  x:=x << 1;
  y++;}
}
\end{lstlisting}
\end{scriptsize}
  \vspace{-10pt} 
\end{wrapfigure} 
Table \ref{tab-ex} presents more comparisons with some existing methods. The first line refers to a program drawn from an industrial audio compression module in~\cite{Cook:2008}, Ex.1 and Fig.3. It is depicted below on the right.  We concentrate on the $\textsf{while}$ loop staring at line.
Following Section \ref{antaffine} and its Theorem \ref{ant-affine}, we have the matrices:
 $A'=\begin{small}\begin{bmatrix}2&0&0\\ 0&1&1\\ 0&0&1 \end{bmatrix}\end{small}$ and $F'=\begin{small}\begin{bmatrix}-1&0&2^{30}\\ 0&0&0\\ 0&0&1 \end{bmatrix}\end{small}$. 
 Matrix $A'$ is already in Jordan normal form with eigenvalues $2$ and $1$ of multiplicity  $1$ and $2$, respectively.  One needs to treat lines $f_1, f_2$ and $f_3$ of  $F'$. Using Theorem \ref{formula} and its ready-to-use formulas \ref{1}, \ref{2} and \ref{3} one computes $ANT(P(A',f_i))$ for $1\leq i\leq 3$. Using the same notations introduced in Section \ref{practice}, we obtain $S_{2,1}\equiv (x_{2,1}<0)$, $S_{1,1}\equiv (x_{2,1}=0)$ and $S_{1,2}\equiv (x_{2,1}=0)$, and the computed $ANT$ set is $S_{2,1}\vee S_{1,1}\vee S_{1,2}$. In other words, in the canonical basis if $u_1$ and $u_2$ are the parameters used as  initial values for the variables $x$ and $y$ we obtain the precise $ANT$ set $(u_1<0)\vee(u_1=0)$.  In \cite{Cook:2008}, the computation of there precondition for termination took $22$ seconds. Our algorithm took $0.03$ seconds to compute our precondition for termination. Moreover, the computed $ANT$ set is exactly the set of non-terminating inputs and thus its complementary set is the exact set of terminating inputs.    
The second line in Table \ref{tab-ex} refers to a program from  
\cite{Bozga} and \cite{Cook:2008}, depicted below.

\begin{wrapfigure}{r}{0.15\textwidth}
  \vspace{-10pt}
    \begin{small}
\begin{lstlisting}
while(x>0){
 x:=x+y;
 y:=y+z;
}
\end{lstlisting}
\end{small}
  \vspace{-10pt}
  \vspace{-10pt}
\end{wrapfigure}
THe $ANT$ complementary set obtained by our prototype gives a more precise preconditions for termination than the one previously proposed. 
Our algorithm took $0.03$ seconds to compute our precondition for termination. Moreover, the computed $ANT$ set is exactly the set of non-terminating inputs and thus its complementary set is the exact set of terminating inputs.  The second line in Table \ref{tab-ex} refers to a program from \cite{Bozga} and \cite{Cook:2008}.
For this program, we generate a more precise precondition, representing a larger set of inputs. 
The experiments  in  \cite{Cook:2008}, involving industrial examples and handwritten programs, indicate that there techniques took $24$ seconds to output there preconditions.  The sixth entry in Table \ref{tab-ex} deals with a simple program with a no existing linear ranking function. Also the eighth example deals with a program terminating on $\Z$ but not on $\Q$. The last line of \ref{tab-ex} shows that our algorithm took $1.028$ seconds to handle $32$ loops taken from \cite{samir:2013}.  
\begin{table}[t]
\caption{$ANT^c$ for linear programs from related works}\label{tab1} 
\begin{center}
\begin{scriptsize}
\begin{tabular}{|c|c|c|c|c|c|c|}
\hline 
 Programs& Class & \#Cond & \#Var & Cpu/s:$ANT^c$\\ 
\hline \cite{Cook:2008} Ex.1 Fig.3 & $P^{\A}$ &  $1$ & $2$ & $0.03$\\ 
\hline \cite{Bozga} and \cite{Cook:2008} Ex.1 Fig.4 & $P^{\H}$ &  $1$ & $3$ & $0.02$\\
\hline \cite{tiwaricav04} Ex.3& $P^{\H}$ &  $1$ & $2$ & $0.02$\\
\hline \cite{Podelski04} and \cite{tiwaricav04} Ex.4& $P^{\G}$ &  $2$ & $2$ & $0.03$\\ 
\hline \cite{tiwaricav04} Ex.5& $P^{\G}$ &  $2$ & $2$ & $0.03$\\ 
\hline \cite{Podelski04} Ex.2& $P^{\A}$ &  $1$ & $1$ & $0.02$\\ 
\hline \cite{Braverman} Ex.2& $P^{\H}$ &  $1$ & $2$ & $0.04$\\
\hline \cite{2013:POPL} (1)& $P^{\A}$ &  $2$ & $2$ & $0.05$\\
\hline \cite{2013:POPL} Ex.4.15& $P^{\A}$ &  $4$ & $3$ & $0.06$\\
\hline \pbox{5cm}{32 loops from \cite{samir:2013}:\\ Tab1.\#10 to \#41} & $P^{\H},P^{\G},P^{\A}$ &  $[1,2]$ & $[1,4]$ & $1.28$\\
\hline
\end{tabular}
\end{scriptsize}
\label{tab-ex}
\end{center}
\end{table}

\section{Discussion}\label{discus}
Concerning \emph{termination analysis} for affine programs over the reals, rationals and the integers, we reduced the problem to the emptiness check of the generated $ANT$ sets.
By so doing, we obtained a characterization of terminating linear programs which allows for a practical and complete polynomial time computational procedure. In~\cite{Braverman, tiwaricav04}, the authors focused on the decidability of the termination 
problem for linear loop programs.  Also \cite{Braverman} is based on the approach in \cite{tiwaricav04}, but now considering termination analysis over the rationals and integers for \emph{homogeneous programs}. But the termination problem for \emph{general affine} programs over the integers is left open in~\cite{Braverman}. Our criteria for termination over stable subspaces allowed us to address this question, and lead to new decidability results.
See Section \ref{stable}. In fact, we show that the termination problem for linear/affine program over the integers with real spectrum is decidable. 
Recently, in \cite{stolen}, considering the $ANT$ set with a technique similar to our approach proposed in \cite{scss2013rebiha, TR-IC-14-09}, the authors were able to answer this question for programs with semi-simple matrices, using strong results from analytic number theory, and diophantine geometry. But, the work of \cite{stolen} focus on a decidability results and the $ANT$ set is not explicitely computed. In fact, they refer to the $ANT$ set as a semi-algebraic set (i.e., it is not proved to be semi-linear in there paper) and suggest the use of quantifier elimination.  In a companion article, we provide the more complete response to this open problem and show the decidability for almost all the class of linear/affine programs over $\Z$ except for an extremely small class of Lesbegue measure zero. Also, the contributions of this article is not restricted to decidability results, we provide efficient computational methods to compute the $ANT$ set allowing new termination and conditional termination analysis.

In this work, although we also considered the termination problem, 
we addressed a more general problem, namely, the \emph{conditional termination} problem of generating static sets of terminating and non-terminating inputs  for the program. 
We provided efficient computational methods allowing for the exact computation and symbolic representation of the $ANT$ sets for affine loop programs over $\R$, $\Q$,  $\Z$, and $\N$.
The $ANT$ sets generated by our approach can be seen as a precise over-approximation
for the set of non-terminating inputs for the program.
Here, we use ``precise'' in the sense that $NT \subseteq ANT$ and all elements in $ANT$, even those  not in $NT$, are directly associated with non-terminating values modulo a finite numbers of loop iterations.  
The, possibly infinite, complement of an $ANT$ set is also a ``precise'' under-approximation of the set of terminating inputs, as it provides terminating input data entering the loop at least once.  
Our methods differs from \cite{Cook:2008} as we do not use the synthesis of ranking functions.  The methods proposed in \cite{Gulwani:2008} can provide non-linear preconditions, but we always generate semi-linear sets as precondition for termination, which facilitates the static analysis of liveness properties. The approaches proposed in \cite{Bozga} considers first octagonal relations 
and suggests the use of quantifier elimination techniques and algorithms, which would running in exponential time complexity $O(n^3\cdot 5^n)$. 
They also consider the conditional termination problem for restricted subclasses of
linear affine relations where the  associated matrix has to be diagonalizable,
and  with all non-zero eigenvalues of multiplicity one. The experiments in \cite{samir:2013}, involving handwritten programs, are handled succesfully by our algorithm. The strength and the practical efficiency of the approach is shown by our
experiments dealing with a large number of larger linear loops. 

Our main results, Theorems \ref{equivalence}, \ref{thmcomplex} , \ref{reduc-H}, \ref{ant-affine}, \ref{K}, \ref{SUV}, \ref{thmE0K}, \ref{formula}  and their associated corollaries and direct encodings, are evidences of the novelty of our approach.  


\section{Conclusion}\label{conclusion}

We presented the new notion of \emph{asymptotically non-terminating initial variable values} for linear programs. Our theoretical results provided us with powerful computational methods allowing for the automated generation of the sets of all asymptotically non-terminating initial variable values, represented symbolically and exactly by a semi-linear space, \emph{e.g.}, conjunctions and disjunctions of linear equalities and inequalities.  
We reduced the termination/non-termination problem of linear, affine programs to the emptiness check of the $ANT$ set of specific 
homogeneous linear programs. Moreover, by taking the complement of the semi-linear set of ANT initial variable values, we obtained a precise under-approximation of the set of terminating initial values for such programs. 

These theoretical contributions are mathematical in nature with proofs that are quite technical. We showed, however, that these results 
can be directly applied in practical ways: one can use the ready-to-use formulas representing the ANT set provided in this article. Any static program analysis could incorporate, by a simple and direct instantiation techniques illustrated in our examples, the generic ready-to-use formulas representing precondition for (non-)termination, which were provided.  
This method was also used to tackle the termination and non-termination problem of linear/affine programs on rational or integer initial values, leading to new decidability results for these classes of programs. 

\bibliographystyle{splncs}
\bibliography{termination}
\appendix
\section{Appendix}

\subsection{Proofs of Section \ref{ANT}}

\begin{proof}\textbf{[Theorem \ref{equivalence}]}
It is clear that if $P(\A,\f)$ is $NT$, it is $ANT$ as a $NT$ value of $P(\A,\f)$ is of course $ANT$ (with $k_\x=0$). Conversely, if $P(\A,\f)$ is $ANT$, call $\x$ an $ANT$ value, then $A^{k_\x}(\x)$ is a $NT$ value of $P(\A,\f)$, so $P(\A,\f)$ is $NT$. The assertion for $K$-stable subspaces of $E$ is 
obvious, the proof being the same, as if $x\in K$ is $ANT$, we have $A^{k_\x}(\x)\in K$.
\end{proof}

\begin{proof}\textbf{[Corollary \ref{co-ANT}]}
As $NT(P(\A,\f)) \subseteq ANT(P(\A,\f))$, passing to complementary sets gives the result. 
\end{proof}     
\subsection{Proofs of Section \ref{complex}}
\begin{proof}\textbf{[Proposition \ref{x+}]}
We write $\x^+=\x_1+\dots+ \x_t$, with $\x_i\in E_{\l_i}(\A)$. 
There are polynomials $P_1,\dots,P_t$ in $\R[T]$, 
such that $\f(\A^k(\x))=\l_1^kP_1(k)+\dots+ \l_t^kP_t(k)$. Let $k_0$ be the smallest integer $i$ such that $P_i$ is nonzero, and set $\l=\l_i$. Then $\f(\A^k(x))$ becomes equivalent for $k$ large, to $a\l^k k^m$ for some $a$ the leading coefficient of $P_{k_0}$, and $m$ the degree of $P_{k_0}$. On the other hand, according to Theorem \ref{nadthm}, the program $P(\A,\f)$ is terminating on 
$E'$, hence on $\x'$, and more generally on any $\A^l(x')$ for $l\geq 0$ because $E'$ is $\A$-stable. In particular, there is an infinity of $l\geq 0$, 
such that $\A^l(x^-)$ is $\leq 0$. This implies that if $P(\A,\f)$ is asymptotically non terminating on $\x=\x^+ +\x'$, then we have $a>0$, and thus $P(\A,\f)$ is asymptotically non terminating on $x^+$. 
\end{proof}
\begin{proof}\textbf{[Theorem \ref{xr}]}
Indeed, if one considers $\A^2$ instead of $\A$, then $\x^r$ is equal to $\x^+$ for $\A^2$, and 
$\x^{nr}$ corresponds to $\x'$ for $\A^2$. As $P(\A,\f)$ is ANT on $\x$, so is $P(\A^2,\f)$. But according 
to the previous proposition, this means that $P(\A^2,\f)$ is ANT on $\x^+$. Similarly, 
$\A(\x^r)=\A(\x)^r$, is equal to $\A(\x)^+$ for $\A^2$, and $\A(\x^{nr})=\A(\x)^{nr}$, is equal to $\A(\x)'$ for $\A^2$. Again, $P(\A^2,\f)$ 
is $ANT$ on $\A(\x)$, hence by the same argument, it is ANT on $\A(\x^r)$. We thus conclude that $P(\A,\f)$ is ANT on $\x^{r}$. 
\end{proof}
\begin{proof}\textbf{[Theorem \ref{thmcomplex}]}
It is clear that if $ANT^r(P(\A,\f))$ is non empty, then $ANT(P(\A,\f))$ is not. The converse is a consequence of Theorem \ref{xr}. 
The last equality is almost by definition.
\end{proof}
\subsection{Proofs of Section \ref{generaltermination}}
\begin{proof}\textbf{[Proposition \ref{crux}]}
Suppose that $P(\A,\F)$ is $ANT$. Then there is $\x$ in $\underset{i}{\bigcap} ANT(P(\A,\f_i))$. 
Now we write $x=x^r+x^{nr}$ in a unique way. According to 
Theorem \ref{xr}, we know that $x^r$ is is $ANT$ for every $P(\A,\f_i)$, hence it belongs to 
$\underset{i}{\bigcap} ANT^r(P(A,\f_i))$ which can't be empty.
\end{proof}
\begin{proof}\textbf{[Theorem \ref{reduc-H}]}
We recall that $P(\A,\F)$ is $NT$ if and only if it is $ANT$ thanks to Lemma \ref{ANTandNT}. The proof then follows 
from Proposition \ref{crux} and the inclusion $ANT^r(P(\A,\F))\subset ANT(P(\A,\F))$.
\end{proof}
\begin{proof}\textbf{[Theorem \ref{ant-affine}]}
To say that $(x, 1)^{\top}$ is $ANT$ for $P(A',F')$ means that there exists $k_x$, such that if $k\geq k_x$, then $F' A'^k \cdot (x,1)^{\top}$ is $>0$. But as 
$A' \cdot (x, 1)^{\top}= \begin{small}\begin{pmatrix} Ax+c \\ 1 \end{pmatrix}\end{small} = (x_1 , 1)^{\top}$, by induction, we obtain $A'^k \cdot (x , 1)^{\top} = (x_k , 1)^{\top}$. Now, we obtain the relation 
$F' A'^k \cdot (x ,1 )^{\top}=F' \cdot (x_k, 1)^{\top} =  \begin{small}\begin{pmatrix} Bx_k-b \\ 1 \end{pmatrix}\end{small}$. Hence 
$F' A'^k \cdot (x , 1)^{\top}>0$ is equivalent to $Fx_k>b$, and the result follows.
\end{proof}
\subsection{Proofs of Section \ref{stable}}
\begin{proof}\textbf{[Proposition \ref{real2real}]}
If $A$ has a real spectrum, then $Spec(A')=Spec(A)\cup \{1\}$ is also real. In particular $ANT(P(A',F'))=ANT^r(P(A',F'))$. The result follows.
\end{proof}
\begin{proof}\textbf{[Theorem \ref{K}]}
We saw that $x\in K$ is ANT (resp. NT) for $P(A,F,b,c)$ if and only if $x'=(x, 1)^{\top}$ is ANT (resp. NT) for $P(A',F')$, with 
$A'=\begin{bmatrix} A & c \\ & 1 \end{bmatrix}$, and $F'=\begin{bmatrix} F &-b \\ 0 & 1\end{bmatrix}$. We then apply Theorem \ref{equivalence}, to the subset 
$K'=\{x',x\in K\}$ of $\R^{n+1}$, which is $A'$-stable.
\end{proof}
\subsection{Proofs of Section \ref{reduced}}
\begin{proof}\textbf{[Lemma \ref{goodbasis}]}
Let $(e_1,\dots,e_{d_\l})$ is a Jordan basis of $E_\l(\A)$, i.e. a basis $J_\l$ (which exists by classical linear algebra) of $E_{\l}(\A)$ such that 
$Mat_{J_\l}(\A)=diag(U_{\l,1},\dots,U_{\l,r_\l})$, where each $U_{\l,i}$ is of the form 
\begin{tiny}$\begin{pmatrix} \l & 1 &  &   &   &  \\
  & \l & 1&   &   &  \\
  &  & \ddots &\ddots &   & \\
  &  &  & \l & 1  & \\
  &  &  &  & \l  & 1 \\
    &  &  &  &   & \l
\end{pmatrix}$\end{tiny}. We simply take $B_\l=(e_1,\l^{-1}e_2,\dots,\l^{1-d_\l}e_{d_\l})$.
\end{proof}
\begin{proof}\textbf{[Proposition \ref{oneblock}]}
If it was not the case, $\A$ would have two linearly independent eigenvectors $v$ and $w$ associated to $\l$. But as $K(\A,\f)$ is zero, 
$\f(\A^k(v))$ is non constantly zero. As it is equal to $\l^k \f(v)$, we obtain $\f(v)\neq 0$, hence we can actually normalize $v$ 
so that $\f(v)=1$. Similarly, we can suppose that $\f(w)=1$. But then, $\f(\A^k(v-w))$ is constantly $0$, i.e. $v-w\in K(\A,\f)$, which contradicts $K(\A,\f)=\{0\}$.
\end{proof}
\begin{proof}\textbf{[Lemma \ref{poweroflambda}]}
For $k=1$, it is by definition of $T_\lambda$. We now do the induction step from $k$ to $k+1$, for $k\geq 1$. We have 
$(T_\lambda)^{k+1}=T_\lambda(T_\lambda)^k $. But multiplying on the left by $T_\lambda$ a matrix $A$, amounts to replace every row of $A$, by 
this row added with the same one. But by Pascal's triangle equality, we have $(_i^k)+(_{i+1}^k)=(_{i+1}^{k+1})$, and this shows the induction step.
\end{proof}
\begin{proof}\textbf{[Proposition \ref{version1}]}
It is a consequence of Lemma \ref{poweroflambda}, as $Mat_B(\f\circ \A^k)=Mat_B(\f)Mat_B(\A^k)$.
\end{proof}
\begin{proof}\textbf{[Proposition \ref{1}]}
As $P_\mu$ is zero as soon as $|\mu|> \l$ we know that, asymptotically, $f(\A^k(x))$ is equivalent to 
$\l^k(P_{|\l|}(x_{|\l|},k)+(-1)^k P_{-{|\l|}}(x_{-|\l|},k))$. So we have that $f(\A^{2k}(x))$ is equivalent to $\l^{2k} Q_{\pm\l}^+(x_{\pm\l},2k)$, and 
$f(\A^{2k+1}(x))$ is equivalent to $\l^{2k+1} Q_{\pm\l}^-(x_{\pm\l},2k+1)$. In particular, for $k$ large enough, both quantities 
$\l^{2k} Q_{\pm\l}^+(x_{\pm\l},2k)$ and $\l^{2k+1} Q_{\pm\l}^-(x_{\pm\l},2k+1)$ will be positive by our assumption on the dominant terms of 
$ Q_{\pm\l}^+$ and $ Q_{\pm\l}^-$. This means that $x$ is $ANT$.
\end{proof}
\begin{proof}\textbf{[Proposition \ref{1'}]}
 One just needs to expand the polynomials $Q_{\pm\l}^+(x_{|\l|})$, $Q_{\pm\l}^-(x_{|\l|})$, and $P_\mu(x_\mu)$. Their coefficients 
are the linear forms $\phi_{\pm\l,i}(x_\l,x_{-\l})$ and $\phi_{\mu,j}(x_\mu)$ involved in the statement. Now we simply express 
the fact that a polynomial is zero if and only if its coefficients are zero, and that
 $Q_{\pm\l}^+(x_{|\l|})$ and $Q_{\pm\l}^-(x_{|\l|})$ have positive dominant coefficient if and only if their 
first coefficients are zero, and the first nonzero one occurring is positive.
\end{proof}
\begin{proof}\textbf{[Proposition \ref{2}]}
Because of our first condition, the quantity $\f(\A^{2k}(x))$ is asymptotically equivalent to $\l^{2k}Q_{\pm\l}^+(x_{\pm\l},2k)$ for $k$ large. Thanks to our second condition, 
the quantity $f(\A^{2k+1}(x)$ is asymptotically equivalent to $|\l'|^{2k+1}Q_{\pm\l'}^-(x_{\pm\l'},2k+1)$. In both cases, as 
$Q_{\pm\l}^+(x_{\pm\l})$ and 
$Q_{\pm\l'}^-(x_{\pm\l'})$ both have positive dominant term, we conclude that $\f(\A^{2k}(x))$ and $\f(\A^{2k+1}(x))$ are both positive when $k$ is large.
\end{proof}
\begin{proof}\textbf{[Proposition \ref{2'}]}
 Similar Proposition \ref{1'}'s proof.
\end{proof}
\begin{proof}\textbf{[Proposition \ref{3}]}
Because of our first condition, the quantity $f(\A^{2k+1}(x))$ is asymptotically equivalent to $\l^{2k+1}Q_{\pm\l}^-(x_{\pm\l},2k+1)$ for $k$ large. 
Thanks to our second condition, $f(\A^{2k}(x))$ is asymptotically equivalent to $|\l'|^{2k}Q_{\pm\l'}^+(x_{\pm\l'},2k)$. 
In both case, as $Q_{\pm\l}^-(x_{\pm\l})$ and 
$Q_{\pm\l'}^+(x_{\pm\l'})$ both have positive dominant term, we conclude that $f(\A^{2k}(x))$ and $f(\A^{2k+1}(x))$ are both positive when $k$ 
is large.
\end{proof}
\begin{proof}\textbf{[Proposition \ref{3'}]}
 Similar to Proposition \ref{1'}'s proof.
\end{proof}
\begin{proof}\textbf{[Theorem \ref{SUV}]}
Let $x$ be an $ANT$ point. If the eigenvalue $\l$ of largest absolute value such that $P_\l(x_\l)$ is nonzero, was negative, and $P_{-\l}(x_{-\l})$ was 
equal to zero, then $\f(\A^k(x)$ would be asymptotically equivalent to $\l^k P_\l(x_\l,k)$. As $P_\l(x_\l,k)$ is asymptotically of the sign of its dominant term, and $\l^k$ is alternatively positive and negative, the program $P(\A,\f)$ would terminate on $x$. Hence, if $\l$ is of largest absolute value 
such that $P_\l(x_\l)$ is nonzero, then $P_{|\l|}(x_{|\l|})$ is nonzero, and we can actually suppose that $\l$ is positive.
If $Q_{\pm\l}^+(x_{\pm\l})$ and $Q_{\pm\l}^-(x_{\pm\l})$ are nonzero, as $\f(\A^{2k}(x)) \sim \l^{2k} Q_{\pm\l}^+(x_{\pm\l},2k)$ and 
$\f(\A^{2k+1}(x)) \sim \l^{2k+1} Q_{\pm\l}^-(x_{\pm\l},2k+1)$, they will both be positive for $k$ large if and only if $Q_{\pm\l}^+(x_{\pm\l})$ and $Q_{\pm\l}^-(x_{\pm\l})$ have a positive dominant term. In this case, we are in the situation of Proposition \ref{1}. If $Q_{\pm\l}^-(x_{\pm\l})$ is equal to zero, then $Q_{\pm\l}^+(x_{\pm\l})$ is not (otherwise $P_{\l}(x_\l)$ and $P_{-\l}(x_{-\l})$ would both 
be zero), so $\f(\A^{2k}(x)) \sim \l^{2k} Q_{\pm \l}^+(x_{\pm\l},2k)$, and $Q_{\pm\l}^+(x_{\pm\l})$ must have a positive dominant term. If 
$Q_{\pm\mu}^-(x_{\pm\mu})$ was zero for all eigenvalue $\mu$, we would have $\f(\A^{2k+1}(x))=0$ for all $k$, which is absurd because $x$ is ANT. Hence there is $\l'$ of absolute value as large as possible (with $|\l'|<\l$ necessarily), such that $Q_{\pm\l'}^-(x_{\pm\l'})$ is nonzero. In this case, $\f(\A^{2k+1}(x))$ is equivalent to $|\l'|^{2k+1}Q_{\pm\l'}^-(x_{\pm\l'})$, and as $x$ is ANT, this forces 
$Q_{\pm\l'}^-(x_{\pm\l'})$ to have a positive dominant term, and we are in the situation of Proposition \ref{2}. Finally, in the last case, $Q_{\pm\l}^+(x_{\pm\l})$ is equal to zero, and $Q_{\pm\l}^-(x_{\pm\l})$ is necessarily 
nonzero. As $\f(\A^{2k+1}(x)) \sim \l^{2k+1} Q_{\pm\l}^-(x_{\pm\l},2k+1)$, this implies that $Q_{\pm\l}^-(x_{\pm\l},2k+1)$ has a positive 
dominant term. Again, if $Q_{\pm\mu}^+(x_{\pm\mu})$ was equal to zero, for all $k\geq 0$, we would have $\f(\A^{2k}(x))=0$, which is absurd as $x$ 
is ANT. Hence there is $\l'$ of absolute value as large as possible (with $|\l'|<\l$ necessarily), such that $Q_{\pm\l'}^+(x_{\pm\l'})$ is nonzero. In this case, $\f(\A^{2k}(x))$ is equivalent to $|\l'|^{2k}Q_{\pm\l'}^+(x_{\pm\l'})$, and as $x$ is ANT, this forces 
$Q_{\pm\l'}^+(x_{\pm\l'})$ to have a positive dominant term, and we are in the situation of Proposition \ref{3}. 
\end{proof}
\subsection{Proofs of Section \ref{real}}
\begin{proof}\textbf{[Proposition \ref{polynomial}]}
The restriction of $\A$ to $E_\l(\A)$ admits a matrix of the form 
$T=\begin{tiny}\begin{pmatrix} \l & \dots & . \\ & \ddots & \vdots \\ & & \l \end{pmatrix}\end{tiny}$ in a Jordan basis of $E_\l(\A)$. It is easy to check, by induction 
on $d_\l$, using the theory of Bernoulli polynomials, that $T^k$ is upper triangular, with diagonal entries equal to $\l^k$, and non diagonal 
nonzero entries of the form $\l^kQ(k)$, for $Q\in \R[X]$ of degree $\leq d_\l$. The result follows.
\end{proof}
\begin{proof}\textbf{[Theorem \ref{reductiontoregular}]}
As $\f(\A^k(x))=\overline{\f}(\overline{\A}^k(\overline{x}))$, it is obvious that $x$ is $ANT$ for $P(\A,\f)$ if and only if 
$\overline{x}$ is $ANT$ for $P(\overline{\A},\overline{\f})$. Now if we write $\overline{x}=\overline{x}_0+\overline{x}^a$, 
with $\overline{x}_0$ in $\overline{E}_0(\overline{\A})$, and $\overline{x}^a$ in $\overline{E}^a$, then for $k$ large, 
we have $\overline{\f}(\overline{\A}^k(\overline{x}))=\overline{\f}(\overline{\A}^k(\overline{x}^a))$. This means that 
$\overline{x}$ is $ANT$ for $P(\overline{\A},\overline{\f})$ if and only if $\overline{x}^a$ is $ANT$ for $P(\overline{\A}^a,\overline{\f}^a)$. This ends the proof.
\end{proof}
\subsection{Proofs of Section \ref{practice}}
\begin{proof}\textbf{[Theorem \ref{Zar}]}
For $A$ in $\mathcal{M}(n,\R)$, let $\tilde{A}$ be the block diagonal matrix $diag(A,-A)$ of $\mathcal{M}(2n,\R)$. The spectrum of 
$\tilde{A}$ is $Spec(A)\cup -Spec(A)$, hence if it contains $2n$ different elements, certainly, $A$ will not admit $\l$ and 
$-\l$ as simultaneous eigenvalues. But $Spec(\tilde{A})$ is of cardinality $2n$ if and only $P(A)=Disc(\chi_{\tilde{A}})\neq 0$, where $Disc$
is the discriminant, and $\chi_{\tilde{A}}$ the characteristic polynomial of $\tilde{A}$. As the map $A\mapsto P(A)$ is polynomial, we see 
that $R(A)=\{A\in \mathcal{M}(n,\R),P(A) \neq 0\}=\{A\in \mathcal{M}(n,\R),|Spec(\tilde{A})|=2n\}$ is a Zariski open subset, obviously non empty (take $A=diag(1,\dots,n)$), hence 
$R(A)\times \R^n$. Now we are going 
to show that the set $R'(A)=\{(A,v)\in \mathcal{M}(n,\R)\times \R^n, K(A,v)=\{0\}\}$ is also a Zariski open and nonempty in $\mathcal{M}(n,\R)$. Write 
$B(A,v)\in \mathcal{M}(n,\R)$ the matrix the rows of which are ${}^t\!v,{}^t\!vA,\dots,{}^t\!vA^{n-1}$. Then $R'(A)=\{(A,v)\in\mathcal{M}(n,\R)\times \R^n, 
det(B(A,v))\neq 0\}$, hence is Zariski open in $\mathcal{M}(n,\R)\times \R^n$. To see that it is non empty, take $v$ the first vector of 
the canonical basis of $\R^n$, and $A$ the permutation matrix representing the cycle $(1,\dots,n)$. Finally, the set $R(A,v)$ we are interested contains 
the non empty Zariski open set $R(A)\times \R^n \cap R'(A)$, which proves the statement.
\end{proof} 
\begin{proof}\textbf{[Theorem \ref{thm-algo}]}
 To begin, one need to recall the following observation: if $P_{\l,j}$ is nonzero, its dominant term is the first nonzero 
$a_{\l,m}$, for $m$ between $1$ and $j$.
Suppose that the $\l$ of largest absolute value such that a coefficient $a_{\l,j}x_{\l,j}$ is nonzero is positive, and that $a_{\l,j}x_{\l,j}$ is positive as well. In this case, $\f(\A^k(x))$ is equivalent to $\l^kP_\l(x_\l,k)$. We then recall that $P_\l(x_\l)$ is equal to $\sum_{j=1}^{d_\l} x_{\l,j} P_{\l,j}$. Finally, thanks to Proposition \ref{version1}, we see that the dominant term of $P_\l(x_\l)$ is equal to $a_{\l,j_0}x_{\l,j_0}$, for the largest $j_0$ such that $a_{\l,j_0}x_{\l,j_0}$ is nonzero, and as it is positive, $x$ is $ANT$.\\
Conversely, if $x$ is $ANT$, all $P_\l(x_\l)$ can't be zero, otherwise $\f(\A^k(x))$ would be constantly zero, which is absurd. 
Let $\l$ be the eigenvalue of largest absolute value, such that $P_{\l}(x_\l)$ is nonzero. Then again, $\f(\A^k(x))$ is equivalent to $\l^kP_\l(x_\l,k)$, and as $P_\l(x_\l,k)$'s dominant term is equal to $a_{\l,j_0}x_{\l,j_0}$, for the largest $j_0$ such that $a_{\l,j_0}x_{\l,j_0}$ is nonzero, $\f(\A^k(x))$ is equivalent to $\l^k a_{\l,j_0}x_{\l,j_0}k^{j_0}$ for $k$ large. As $x$ is $ANT$, $\l$ must be positive, because otherwise $\f(\A^k(x))$ would alternatively change sign for $k$ large. Moreover, $a_{\l,j_0}x_{\l,j_0}$ must be positive.
\end{proof}
\end{document}